\documentclass[aps,pra,preprint,groupedaddress]{revtex4-1}
\pdfoutput=1

\usepackage{amsmath}
\usepackage{graphicx}
\usepackage{multirow}
\usepackage[table]{xcolor}
\usepackage[caption=false,font=footnotesize]{subfig}
\newcommand{\etal}{{\em et al.}}
\newcommand{\incfig}{\centering\includegraphics}
\newcommand{\abs}[1]{\left| #1 \right|} % for absolute value
\newcommand{\pd}[2]{\dfrac{\partial #1}{\partial #2}}
\let\baraccent=\= % rename built-in command \= to \baraccent
\renewcommand{\=}[1]{\stackrel{#1}{=}} % for putting numbers above =

\begin{document}

% number in the upper right hand corner of the title page in preprint mode.
%\preprint{}

\title{Numerical Investigation of the Effect of Airfoil Thickness on Onset of
Dynamic Stall}


\author{Anupam Sharma}
\email[]{sharma@iastate.edu}
\thanks{Assistant Professor,}
\affiliation{Department of Aerospace Engineering, Iowa State University, Ames, IA, USA, 50011.}

\author{Miguel Visbal}
\email[]{miguel.visbal@us.af.mil}
\thanks{CFD Technical Advisor,}
\affiliation{Computational Sciences Center, Aerospace Systems Directorate, Air Force Research Laboratory, Wright-Patterson AFB, OH 45433.}


\date{\today}

%\linenumbers
\begin{abstract}
Effect of airfoil thickness on onset of dynamic stall is investigated using
large eddy simulations at chord-based Reynolds number of 200,000. Four
symmetric NACA airfoils of thickness-to-chord ratios of 9\%, 12\%, 15\%, and
18\% are studied. The 3-D Navier Stokes solver, FDL3DI is used with a
sixth-order compact finite difference scheme for spatial discretization,
second-order implicit time integration, and discriminating filters to remove
unresolved wavenumbers. A constant-rate pitch-up maneuver is studied with the
pitching axis located at the airfoil quarter chord point. Simulations are
performed in two steps. In the first step, the airfoil is kept static at a
prescribed angle of attack ($=4^\circ$). In the second step, a ramp function is
used to smoothly increase the pitch rate from zero to the selected value and
then the pitch rate is held constant until the angle of attack goes past the
lift stall point. Comparisons against XFOIL for the static simulations show
good agreement in predicting the transition location. FDL3DI predicts two-stage
transition for thin airfoils (9\% and 12\%), which is not observed in the XFOIL
results. The dynamic simulations show that the onset of dynamic stall is marked
by the bursting of the laminar separation bubble (LSB) in all cases.  However,
for the thickest airfoil tested, the reverse flow region spreads over most of
the airfoil and reaches the LSB location immediately before the LSB bursts and
dynamic stall begins, suggesting that stall could be triggered by the separated
turbulent boundary layer. The results suggest that the boundary between
different classifications of dynamic stall, particularly leading edge stall
versus trailing edge stall are blurred. The dynamic stall onset mechanism
changes gradually from one to the other with a gradual change in some
parameters, in this case, airfoil thickness.
\end{abstract}

%\keywords{}

\maketitle

% --- sections ----
%%%%%%%%%%%%%%%%%%%%%%%%%%%%%%%%%%%%%%%%%%%
\section{Introduction}
\label{sec:intro}
%%%%%%%%%%%%%%%%%%%%%%%%%%%%%%%%%%%%%%%%%%%

Unsteady flow over streamlined surfaces produces interesting but usually
undesirable phenomena such as flutter, buffeting, gust response, and dynamic
stall~\cite{McCroskey_1982}. Dynamic stall is a nonlinear fluid dynamics
phenomenon that occurs frequently on rapidly maneuvering
aircraft~\cite{brandon_1991}, helicopter rotors~\cite{ham_1968}, and wind
turbines~\cite{fujisawa_2001,larsen_2007}, and is characterized by large
increases in lift, drag, and pitching moment far beyond the corresponding
static stall values.  Carr~\cite{Carr_1988} presents an excellent review on
dynamic stall. Dynamic stall can be divided into two categories based on the
degree to which the angle of attack, $\alpha$ increases beyond the static-stall
value. Denoting the maximum $\alpha$ reached during the unsteady motion by
$\alpha_{max}$, these categories are: (1) {\em Light stall:} when
$\alpha_{max}$ is small, the viscous, separated flow region is small (of the
order of the airfoil thickness), and (2) Deep stall: for large $\alpha_{max}$,
the viscous region becomes comparable to the airfoil chord. A prominent feature
of deep stall is the presence of the dynamic stall vortex (DSV) that is
primarily responsible for the large overshoots in aerodynamic forces and
moments. 

Many fundamental aspects of flutter, buffeting, and gust response can be
explained using linearized theory. Pioneering work in this area was done by
Theodorsen~\cite{theodorsen_1935} and Karman and Sears~\cite{karman_1938}. The
linearized approach however is limited to small perturbations and the highly
nonlinear phenomenon of dynamic stall requires other approaches. Semi-empirical
methods~\cite{leishman_1989,ericsson_1988} have been developed to model dynamic
stall. These methods are invaluable for preliminary design and analysis, but
they do not provide insight into the physical mechanisms.  Computational
investigations have included Reynolds Averaged Navier-Stokes (RANS)
computations~\cite{visbal_1990} and large eddy simulations
(LES)~\cite{garmann_2011,visbal_2011}. Recent computational efforts have
focused on using highly resolved LES to investigate dynamic stall on flat
plates~\cite{garmann_2011} and airfoils~\cite{visbal_1990}. All of these
simulations have focused on relatively thin airfoils operating at
low-to-moderate chord-based Reynolds numbers, $10^4 < Re_c < 5\times10^5$. In
this paper, we explore the effects of airfoil geometry on the onset of dynamic
stall at $Re_c=2\times 10^5$ using large eddy simulations. In particular, we
focus on the mechanism of stall onset as airfoil thickness is varied.

%%%%%%%%%%%%%%%%%%%%%%%%%%%%%%%%%%%%%%%%%%%
\section{Methodology}
\label{sec:methodology}
%%%%%%%%%%%%%%%%%%%%%%%%%%%%%%%%%%%%%%%%%%%
The extensively validated compressible Navier-Stokes solver,
FDL3DI~\cite{visbal_2002} is used for the fluid flow simulations. FDL3DI solves
the full, unfiltered Navier-Stokes equations on curvilinear meshes. The solver
can work with multi-block Overset (Chimera) meshes with high order
interpolation methods that extend the spectral-like accuracy of the solver to
complex geometries. The solver can be run in a large eddy simulation (LES) mode
with the effect of sub-grid scale stresses modeled implicitly via spatial
filtering to remove the energy at the unresolved scales. {\em Discriminating},
high-order, low-pass spatial filters are implemented that regularize the
procedure without excessive dissipation. 

%%%%%%%%%%%%%%%%%%%%%%%%%%%%%%%%%%%%%%%%%%%
\subsection{Governing Equations}
\label{sec:geqs}
%%%%%%%%%%%%%%%%%%%%%%%%%%%%%%%%%%%%%%%%%%%
The governing fluid flow equations (solved by FDL3DI), after performing a
time-invariant curvilinear coordinate transform from physical coordinates
$(x,y,z,t)$ to computational coordinates $(\xi, \eta, \zeta, \tau)$, are
written in a strong conservation form as
\begin{equation}
  \pd{}{t}\left( \frac{\bf Q}{J} \right) + \pd{\hat{\bf F}_I}{\xi} + \pd{\hat{\bf G}_I}{\eta} + \pd{\hat{\bf H}_I}{\zeta} 
  = \frac{1}{Re} 
  \left[ 
    \pd{\hat{\bf F}_v}{\xi} + \pd{\hat{\bf G}_v}{\eta} + \pd{\hat{\bf H}_v}{\zeta}
  \right],
  \label{eq:geq_FDL}
\end{equation}
where $J=\partial(\xi,\eta,\zeta,\tau)/\partial(x,y,z,t)$ is the Jacobian of
the coordinate transformation, ${\bf Q}= \{\rho, \rho u, \rho v, \rho w, \rho
E\}$; the inviscid flux terms, $\hat{\bf F}_I,\hat{\bf G}_I,\hat{\bf H}_I$ are
\begin{equation}
  \hat{\bf F}_I  = 
      \begin{bmatrix}
        \rho \hat{U} \\
        \rho u \hat{U} + \hat{\xi}_x p  \\
        \rho v \hat{U} + \hat{\xi}_y p  \\
        \rho w \hat{U} + \hat{\xi}_z p  \\
        (\rho E + p) \hat{U} - \hat{\xi}_t p \\
      \end{bmatrix}, \;
  \hat{\bf G}_I = 
      \begin{bmatrix}
        \rho \hat{V} \\
        \rho v \hat{V} + \hat{\eta}_x p  \\
        \rho v \hat{V} + \hat{\eta}_y p  \\
        \rho w \hat{V} + \hat{\eta}_z p  \\
        (\rho E + p) \hat{V} - \hat{\eta}_t p \\
      \end{bmatrix}, ~{\rm and} \;
  \hat{\bf H}_I = 
      \begin{bmatrix}
        \rho \hat{W} \\
        \rho u \hat{W} + \hat{\zeta}_x p  \\
        \rho v \hat{W} + \hat{\zeta}_y p  \\
        \rho w \hat{W} + \hat{\zeta}_z p  \\
        (\rho E + p) \hat{W} - \hat{\zeta}_t p \\
      \end{bmatrix},
      \label{eq:inviscidFluxes}
\end{equation}
where, 
\begin{align}
  \hat{U} &= \hat{\xi}_t   + \hat{\xi}_x u   + \hat{\xi}_y v   + \hat{\xi}_z w,             \nonumber \\
  \hat{V} &= \hat{\eta}_t  + \hat{\eta}_x u  + \hat{\eta}_y v  + \hat{\eta}_z w,            \nonumber \\
  \hat{W} &= \hat{\zeta}_t + \hat{\zeta}_x u + \hat{\zeta}_y v + \hat{\zeta}_z w,~{\rm and} \nonumber \\
  \rho E  &= \frac{p}{\gamma-1} + \frac{1}{2}\,\rho \,(u^2+v^2+w^2).
\end{align}
In the above, $\hat{\xi}_{(x,y,z)} = J^{-1} \partial \xi/\partial {(x,y,z)}$,
and $u,v,w$ are the components of the velocity vector in Cartesian coordinates,
and $\rho,p,T$ are respectively the fluid density, pressure, and temperature.
The gas is assumed to be perfect, $p = \rho T/\gamma M^2_\infty$. The viscous
flux terms, $\hat{\bf F}_v,\hat{\bf G}_v,\hat{\bf H}_v$ are provided in
Ref.~\cite{visbal_2002a}.

%%%%%%%%%%%%%%%%%%%%%%%%%%%%%%%%%%%%%%%%%%%%
%\subsection{Sound Propagation}
%\label{sec:noise_propagation}
%%%%%%%%%%%%%%%%%%%%%%%%%%%%%%%%%%%%%%%%%%%%
%
%Far-field sound propagation is performed using the Ffowcs Williams-Hawkings
%(FW-H) acoustic analogy~\cite{williams_1969}. By neglecting volume sources
%(non-negligible only at very high flow speeds), the following integral equation
%is obtained for far-field acoustic pressure, $p'$ at location ${\bf x}$ and
%time $t$:
%%
%\begin{equation}
%  p'({\bm x},t) = \frac{1}{4 \pi \abs{1-M_r} \abs{{\bm x}}} 
%  \left[
%    \pd{}{t} \iint [\rho_0 u_i n_i + \rho'(u_i - U_i)n_i] {\rm d} \Sigma +
%    \frac{x_i}{c\abs{{\bm x}}} 
%    \pd{}{t} \iint [p'n_i + \rho u_i(u_j-U_j)n_j] {\rm d} \Sigma
%  \right]
%  \label{eq:fwh}
%\end{equation}
%%
%Solving Eq.~\ref{eq:fwh} requires integrating over a surface $\Sigma$ that
%encloses all sound sources. In the above, $n_i$ is normal to the surface
%$\Sigma$, $p'$ and $\rho'$ are pressure and density fluctuations, $\rho_0$ is
%mean density, $u'_i$ is perturbation flow velocity and $U_i$ is the velocity of
%the surface $\Sigma$. The source is at the origin, and ${\bm x}$ denotes the
%observer location. We choose a ``porous'' surface around the airfoil defined by
%one of the gridlines ($\xi=$ constant $>1$; $\xi=1$ is the airfoil surface) of
%the innermost grid block. The FW-H solver has been validated previously against
%canonical problems (point monopole, dipole, and quadrupole) as well as against
%experimental data for aerodynamic noise from propellers~\cite{sharma_2013}.

%%%%%%%%%%%%%%%%%%%%%%%%%%%%%%%%%%%%%%%%%%%
\subsection{Numerical Scheme}
\label{sec:numerics}
%%%%%%%%%%%%%%%%%%%%%%%%%%%%%%%%%%%%%%%%%%%
%
Finite differencing is used to discretize the governing equations. Space is
discretized using high-order (up to sixth order) compact difference
schemes~\cite{lele_1992}. Time integration is performed using an implicit,
approximately-factored procedure described in Ref.~\cite{visbal_2002}. Spatial
derivatives of any scalar, $\phi$ are obtained in the computational space
($\xi,\eta,\zeta$) by solving the tri-diagonal system -
\begin{equation}
   \alpha \phi'_{i-1} + \phi'_i + \alpha \phi'_{i+1} 
 = \beta \frac{\phi_{i+2}-\phi_{i-2}}{4} + \gamma \frac{\phi_{i+1}-\phi_{i-1}}{2}.
 \label{eq:spderiv}
\end{equation}
Spatial derivatives of different orders of accuracy can be obtained by choosing
different combinations of $\alpha, \beta$, and $\gamma$.  A sixth-order scheme,
obtained by setting $\alpha=1/3,~\gamma=14/9,$ and $\beta=1/9$, is used in this
paper. Equation~\ref{eq:spderiv} is a central scheme which works in the
interior of the domain; for points near the physical and inter-processor
boundaries, one-sided differences are used. Neumann boundary conditions, such
as $\partial p/\partial n=0$, are implemented using fourth-order one-sided
differences. Inviscid fluxes are computed at the node points using
Eq.~\ref{eq:spderiv}. Viscous terms are computed by differentiating the
primitive variables, constructing the viscous flux terms, and then
differentiating the flux terms using Eq.~\ref{eq:spderiv} at the node points.

Since the grid is designed to resolve large, energy-containing eddies (and not
for Direct Numerical Simulations), the content not resolved by the grid (high
wavenumbers) has to be removed from the solution. In traditional LES, this is
achieved via sub-grid scale (SGS) models. In the current simulations, this
objective is achieved by filtering the solution at every sub-iteration during
time integration using the following low-pass, high-order filtering procedure.
Denoting a component of the solution vector (a conserved flow variable) by
$\phi$, its filtered value, $\hat{\phi}$ is obtained by solving the following
system of equations
\begin{equation}
  \alpha_f \hat{\phi}_{i-1} + \hat{\phi}_i + \alpha_f \hat{\phi}_{i+1} 
  = \sum_{n=0}^{N} \frac{a_n}{2}\left( \phi_{i+n} + \phi_{i-n} \right),
  \label{eq:filter}
\end{equation}
where a proper choice of the coefficients, $a_n$ as functions of $\alpha_f$,
with $n$ ranging from $1$ to $N$, results in a $2N^{th}$-order accurate
filtering scheme with a $2N+1$-size stencil. $\alpha_f$ is a free variable that
provides additional control on the degree of filtering achieved for a given
order.  Similar to the implementation of spatial derivatives, one-sided
filtering formulae are used near the boundaries. While the central scheme of
Eq.~\ref{eq:filter} is always dissipative, care needs to be exercised with
one-sided filtering formulae as these can amplify certain wave numbers and make
the solution unstable. In the current simulations, an $8^{th}$-order filter
with $\alpha_f=0.4$ is used in the interior points.

%%%%%%%%%%%%%%%%%%%%%%%%%%%%%%%%%%%%%%%%%%%
\section{Meshing}
\label{sec:mesh}
%%%%%%%%%%%%%%%%%%%%%%%%%%%%%%%%%%%%%%%%%%%

\begin{figure}[htb!]
  \centering
  \incfig[width=\columnwidth]{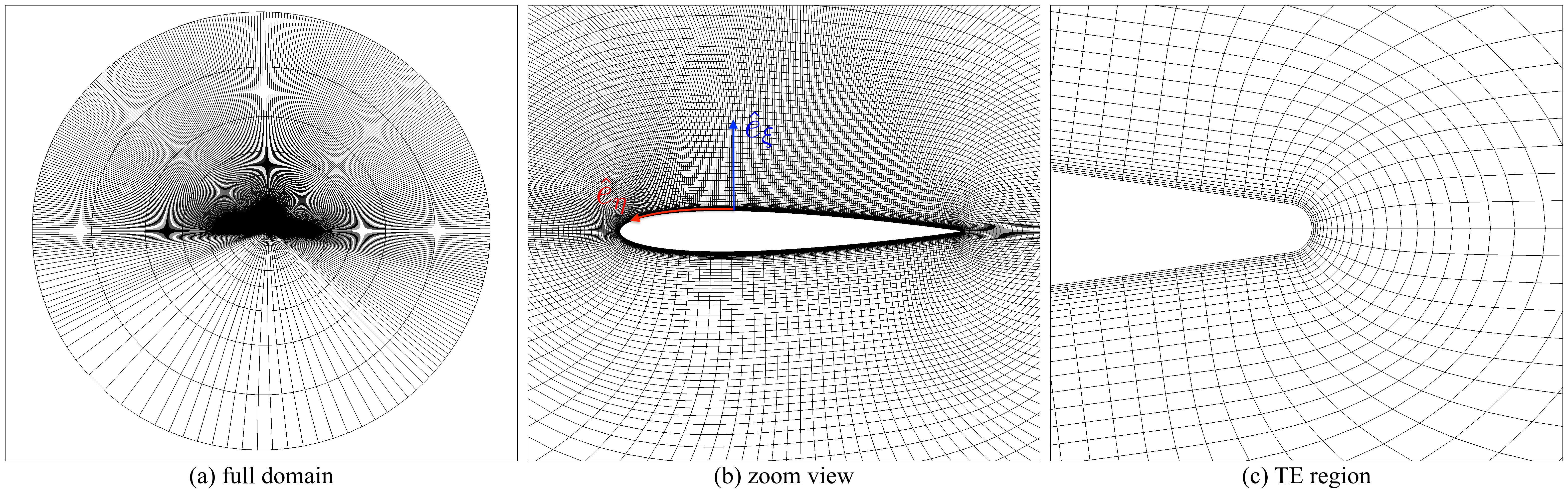}
  \caption{Three views of the mesh used for the NACA 0012 simulation: (a) full
    computational domain, (b) zoom view of the grid around the airfoil, and (c)
    zoom view showing the trailing edge geometry and resolution. Every fifth-
    and every fourth point in the radial and circumferential
    directions respectively are shown for clarity.}
  \label{fig:mesh}
\end{figure}

The simulations are carried out at a chord-based Reynolds number, $Re_c =
200,000$ and a flow Mach number, $M_\infty = 0.1$. The span length of the
airfoil model in the simulations is 10\% of the airfoil chord.
%The first cell height is selected to obtain a $y^+\approx 0.25$. 
A planar, single-block O-mesh is generated around the airfoil, which is
repeated with uniform grid spacing in the span direction. The mesh is highly
refined over the suction side to resolve the viscous flow phenomena expected
during the airfoil pitch up motion.  Figure~\ref{fig:mesh} shows three
cross-sectional views of the computational mesh for one of the airfoils. The
boundary layer on the pressure side stays laminar and attached through most of
the pitch-up maneuver. A relatively coarse mesh is therefore sufficient to
discretize the pressure side. Besides, the dynamic stall phenomenon is
relatively unaffected by the pressure-side flow in the pitch-up maneuver
considered in this study.

The O-grid in the physical space ($x,y,z$) maps to an H-grid in the
computational domain ($\xi,\eta,\zeta$). The following orientation is used:
$\hat{\bm e}_{\xi}$ points radially out, $\hat{\bm e}_{\eta}$ is in the
circumferential direction. Figure~\ref{fig:mesh} (b) shows the orientation of
$\hat{e}_{\xi}$ and $\hat{e}_{\eta}$; $\hat{\bm e}_{\zeta}$ is along the span
direction such that the right hand rule, $\hat{\bm e}_{\zeta} = \hat{\bm
e}_{\xi} \times \hat{\bm e}_{\eta}$ is obeyed. 

%The cell counts in radial,
%circumferential, and spanwise directions are 410, 1341, and 134 respectively
%giving a total cell count of about 74 million. 

Periodic boundary conditions on the $\eta$ boundaries simulate the continuity
in the physical space around the airfoil. Periodicity is also imposed at the
boundaries in the span direction ($\hat{\bm e}_{\zeta}$). Periodic boundary
conditions are implemented using the Overset grid approach in FDL3DI. A minimum
of five-point overlap is required by FDL3DI to ensure high-order accurate
interpolation between individual meshes. A five-point overlap is therefore
built into the mesh. Similar overlaps are created automatically in FDL3DI
between sub-blocks when domain decomposition is used to split each block into
multiple sub-blocks for parallel execution. The airfoil surface is a no-slip
wall. Freestream conditions are prescribed at the outer boundary which is about
100 chords away from the airfoil. The filtering procedure removes all
perturbations as the mesh becomes coarse away from the airfoil to the farfield
boundary.

The same distribution of points around the airfoil is used for the four
airfoils simulated. The same stretching ratios are used to extrude the airfoil
surface grid (along the surface normal direction) to obtain a 2-D O-grid. This
grid is then repeated in the span direction to obtain the final 3-D grid for
each airfoil.

A detailed mesh sensitivity for a constant-rate pitching airfoil has been
presented by Visbal and Garmann~\cite{visbal_2017}. and hence is not repeated
here. The meshes used in the simulations presented here correspond to the
``Fine'' mesh of Ref.~\cite{visbal_2017} with the grid dimensions and first
cell size in wall units ($\Delta x^+, \Delta y^+, \Delta z^+$) presented in
Table~\ref{tab:mesh}.
\begin{table}[htb!]
  \centering
  \caption{Grid dimensions and non-dimensional cell sizes in wall units.
  Averages and max values are over the entire airfoil; the suction side of the
  airfoil is more refined than the pressure side.}
  \label{tab:mesh}
  \begin{tabular}{l|c|c|c|c}
    \hline \hline
    Grid & $N_\xi \times N_\eta \times N_\zeta$ & $\Delta y^+$ (avg, max) & $\Delta x^+$ (avg, max) & $\Delta z^+$ (avg, max) \\
    \hline \hline
%    Coarse & $395 \times 643  \times 51$  &  $0.36,~0.94$  &  $22.0,~55.1$  &  $17.0,~63.9$ \\
%    Medium & $410 \times 995  \times 101$ &  $0.18,~0.47$  &  $14.3,~85.2$  &  $8.5,~32.4$ \\
    Fine   & $410 \times 1341 \times 134$ &  $0.19,~0.50$  &  $10.6,~87.2$  &  $7.0,~24.5$ \\
    \hline \hline
  \end{tabular}
\end{table}

%%%%%%%%%%%%%%%%%%%%%%%%
\section{Results}
\label{sec:results}
%%%%%%%%%%%%%%%%%%%%%%%%
%
The simulations are performed in two steps. In the first step, a statistically
stationary solution is obtained with the airfoil set at $\alpha= 4^\circ$. A
positive $\alpha$ is selected to ensure that the boundary layer on the bottom
surface (pressure side) stays laminar. Dynamic simulations with airfoil motion
are simulated in the second step. A constant-rate pitch-up motion is simulated
with the pitching axis located at the quarter-chord point of the airfoil.
Results of `static' simulations for all three airfoils are presented first.

%%%%%%%%%%%%%%%%%%%%%%%%
\subsection{Static Simulations}
\label{sec:static_results}
\begin{figure}[htb!]
  \incfig[width=0.35\columnwidth]{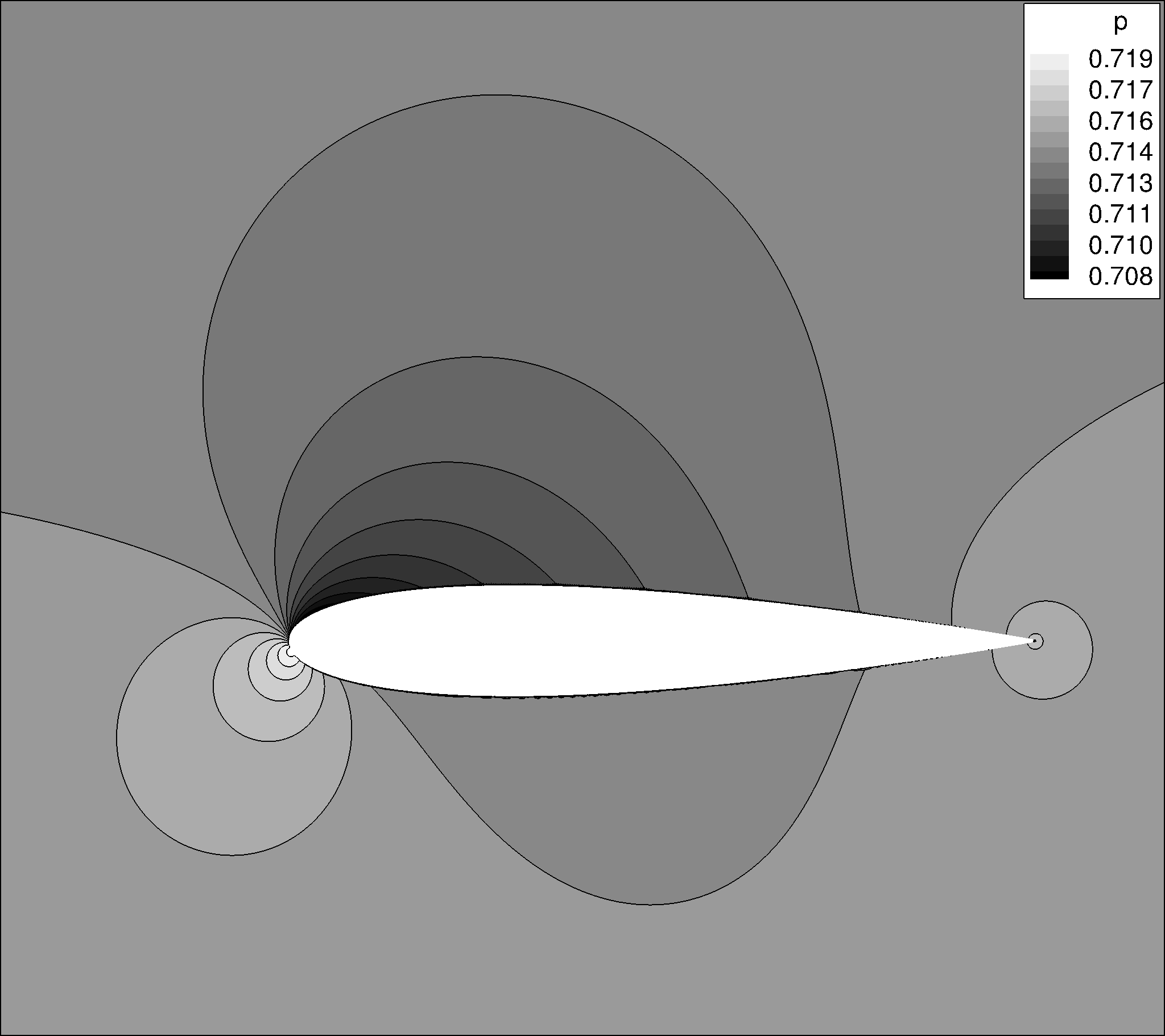}
  \caption{Potential flowfield (NACA-0015 case shown) used to initialize the
    static 2-D viscous simulations.}
  \label{fig:panelInitialization}
\end{figure}

%%%%%%%%%%%%%%%%%%%%%%%%
%
For the static simulations, the $x$ axis of the coordinate system is aligned
with the airfoil chord and constant inflow is prescribed at the desired angle
of attack ($\alpha=4^\circ$ here). In order to minimize the computation time, a 2-D
viscous solution is first obtained by removing the span dimension.  The
two-dimensional solution is computed on a grid that is reduced in the span
direction to three cells, which is the minimum required by FDL3DI to compute an
effectively 2-D solution. Potential flowfield, obtained using an in-house
vortex panel code, is prescribed as the initial condition for the 2D viscous
simulation (see Fig.~\ref{fig:panelInitialization}). The potential solution
sets the pressure and velocity distribution in the farfield to be reasonably
close to the final viscous solution, and avoids large pressure waves that would
otherwise develop if a uniform flowfield is prescribed as the initial
solution. The 2D simulation is run until integrated aerodynamic lift and drag
forces converge.

Static, three-dimensional simulations are then performed with the 2-D viscous
solution repeated in span to generate the initial solution. The simulation is
run until statistical convergence is reached for integrated airfoil loads, as
well as for static pressure at a few point probes placed in the suction side
boundary layer. Surface properties, such as aerodynamic pressure coefficient
($C_P$) and skin friction coefficient ($C_f$) are extracted and compared
against XFOIL predictions. XFOIL~\cite{xfoil} is a panel method code that
simultaneously solves potential flow equations with boundary integral
equations. It uses the $e^{N}$-type amplification formulation to determine
boundary layer transition.

\begin{figure}[htb!]
  \incfig[width=\columnwidth]{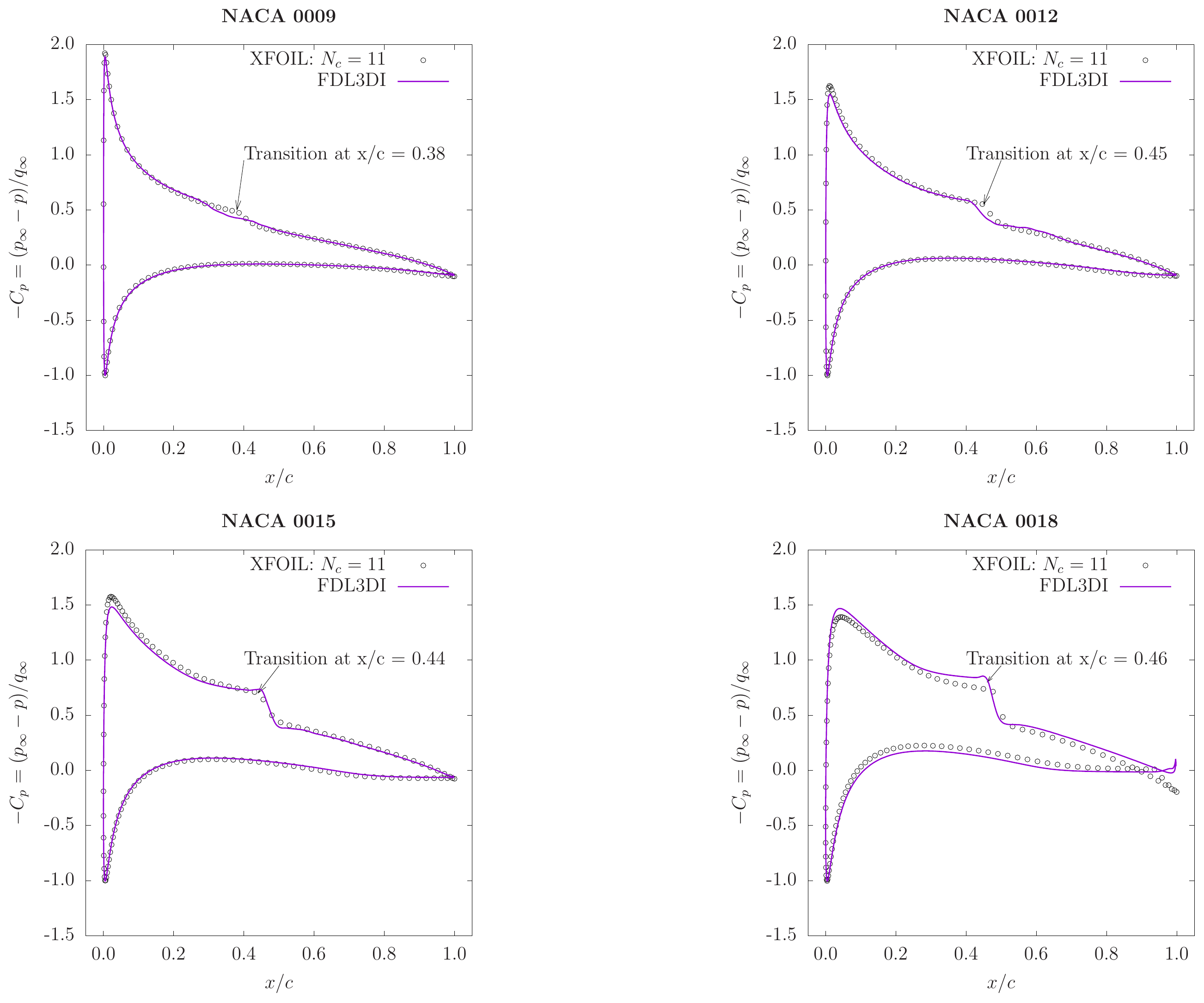}
  \caption{Comparison of coefficient of pressure, $C_P$ between predictions by
    FDL3DI and XFOIL. XFOIL is run with $N_{crit}=11$ to simulate very low
    inflow turbulence.}
  \label{fig:Cp}
\end{figure}

\begin{figure}[htb!]
  \incfig[width=\columnwidth]{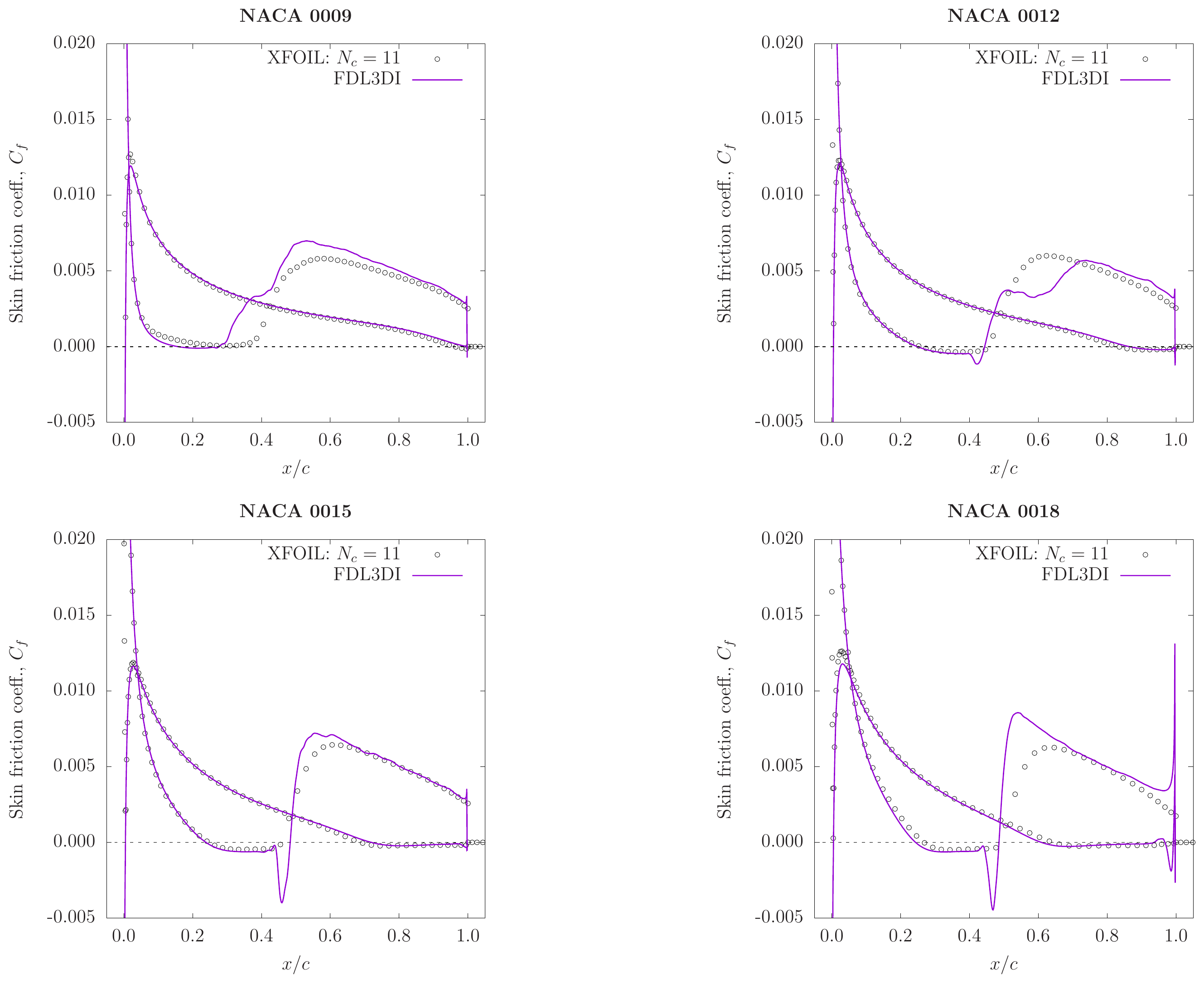}
  \caption{Comparison of skin friction coefficient, $C_f$ between predictions
    by FDL3DI and XFOIL ($N_{crit}=11$).}
  \label{fig:Cf}
\end{figure}

Figures~\ref{fig:Cp} and~\ref{fig:Cf} compare the FDL3DI predicted $C_P$ and
$C_f$ distributions against those obtained using XFOIL for the four airfoils.
The XFOIL simulations are performed with the $N_{crit}$ parameter set equal to
11. $N_{crit}$ is the log of the amplification factor of the most-amplified
wave that triggers transition. A value of 11 for $N_{crit}$ is appropriate for
use with airfoil models tested in a ``clean'' wind tunnel (i.e., with very low
inflow turbulence). Since the inflow in FDL3DI simulations is uniform with zero
turbulence, $N_{crit}=11$ is deemed appropriate.

The overall agreement between XFOIL and FDL3DI is good; the similarities and
the differences are identified here with their possible causes. The peak
suction pressure predictions by the two codes are in good agreement. Highest
peak suction pressure is observed for the thinnest (NACA-0009) airfoil due to
the smallest radius of curvature and the correspondingly high local
acceleration. The transition location can be identified by a sudden drop in
suction pressure; this drop is subtle, especially for the NACA-0009 airfoil.
Transition location is identified more readily with a sudden increase in $C_f$
as seen for all four airfoils in Fig.~\ref{fig:Cf}. Both methods predict nearly
the same location for transition; the largest mismatch is for the NACA-0009
airfoil.  FDL3DI predicts a longer transition region than XFOIL - the $C_f$
curve rises abruptly (a little earlier than XFOIL) marking transition, then
plateaus, and then rises again to its local peak value corresponding to a fully
turbulent boundary layer. A similar, ``two-stage'' transition is seen in FDL3DI
prediction for the NACA-0012 airfoil as well. Similar behavior has been
observed by Barnes and Visbal~\cite{barnes2016}. XFOIL simulations do not
exhibit this two-stage transition, likely because of the simple transition
model, which ensures a monotonic increase in $C_f$ once transition is
triggered. FDL3DI simulations show a large difference between airfoils in $C_f$
distribution around the transition location - the thicker airfoils show a very
steep spatial gradient in chordwise direction ($\partial C_f/\partial x$)
compared to the thin airfoils. This behavior is not predicted by XFOIL, which
shows almost no change in $\partial C_f/\partial x$ with airfoil thickness. In
all the cases simulated here, the laminar boundary layer separates ($C_f<0$),
transition occurs in the shear layer, and the turbulent boundary layer then
reattaches to the surface.

\begin{figure}[htb!]
  \subfloat[NACA-0009]{\incfig[width=0.9\columnwidth]{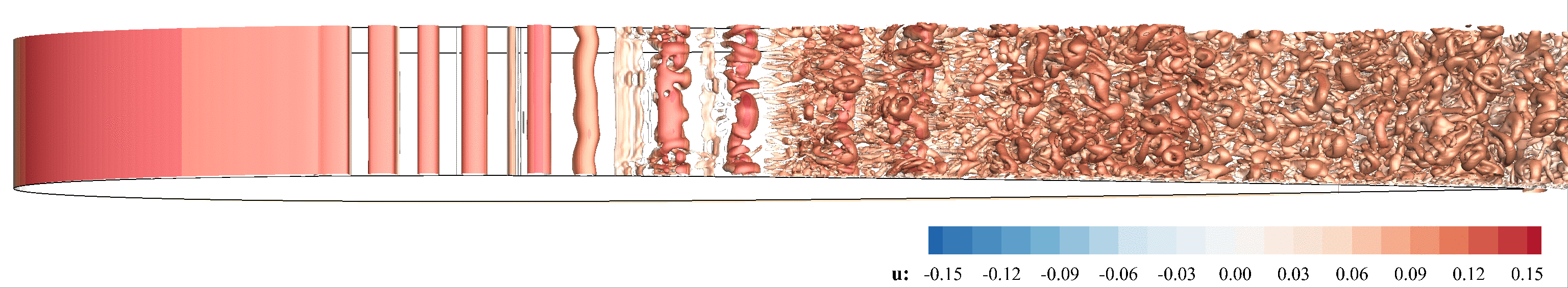}} \\
  \subfloat[NACA-0012]{\incfig[width=0.9\columnwidth]{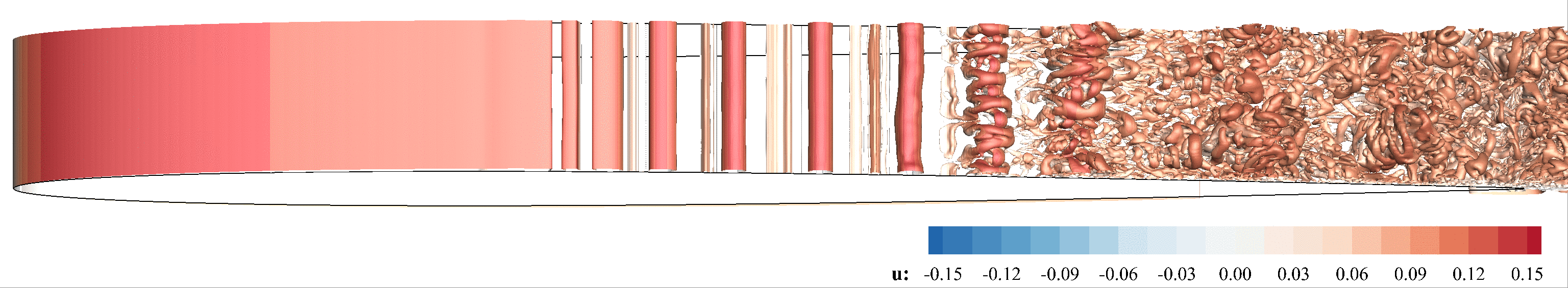}} \\
  \subfloat[NACA-0015]{\incfig[width=0.9\columnwidth]{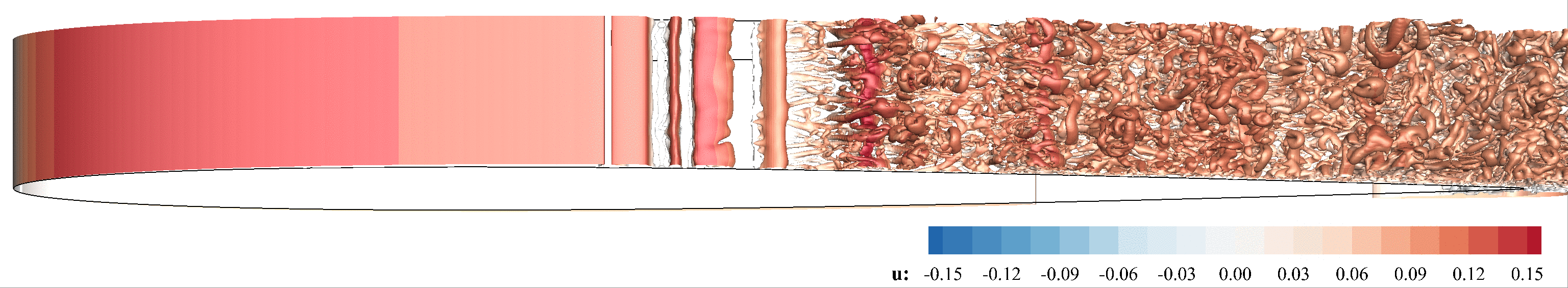}} \\
  \subfloat[NACA-0018]{\incfig[width=0.9\columnwidth]{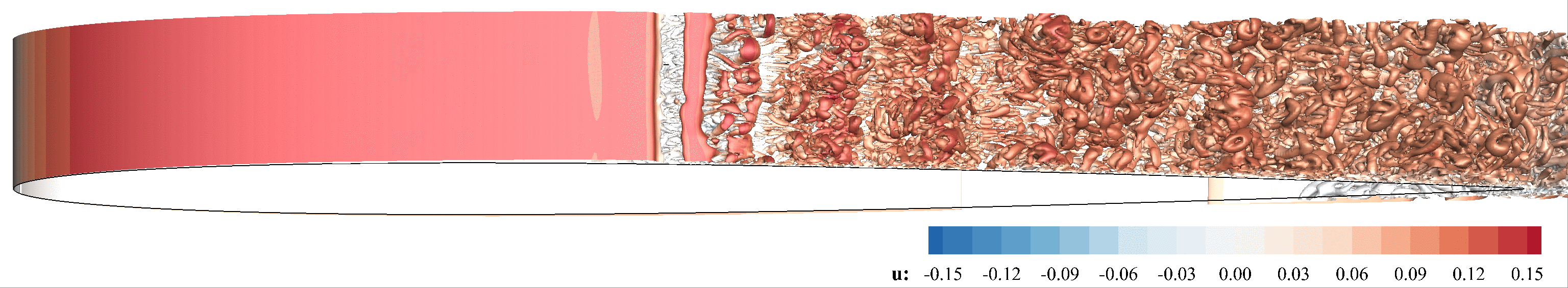}} 
  \caption{Iso-surfaces of Q-criterion to visualize vortical structures near
    the transition region. The transitional region is long for thin airfoils and
    short for thick airfoils.}
  \label{fig:transition_structures}
\end{figure}

To investigate the two-stage transition observed in FDL3DI simulations for
NACA-0009 and NACA-0012, the flow structure near transition location is
investigated. Figure~\ref{fig:transition_structures} shows iso-surfaces of
Q-criterion colored by contours of streamwise velocity. The spanwise coherent
2-D vortex structures (seen clearly for NACA-0009 and NACA-0012) are the
instability waves that breakdown and transition the boundary layer to
turbulence. It is apparent from the figure that the transition region is much
longer for NACA-0009 and NACA-0012 airfoils, while transition occurs over a
much smaller region for NACA-0015 and NACA-0018 airfoils. The long transition
region for the relatively thinner airfoils is the reason why the time averaged
$C_f$ distributions show a two-stage transition, with the plateau representing
the region where the boundary layer is transitional. The higher adverse
pressure gradients in the aft portion of the thicker airfoils is possibly the
reason why the flow breaks down faster and transition occurs abruptly for these
airfoils.

%%%%%%%%%%%%%%%%%%%%%%%%
\subsection{Dynamic Simulations}
\label{sec:dynamic_results}
%%%%%%%%%%%%%%%%%%%%%%%%
%
In the second step, the airfoil pitch-up motion is simulated via grid motion.
A constant-pitch rate motion, with the pitching axis located at the airfoil
quarter-chord point, is investigated. The non-dimensional rotation (pitch) rate
is $\Omega^+_0 = \Omega_0 c/u_\infty = 0.05$.  An abrupt change of rotation
rate from zero to a finite value would result in a very large acceleration
(limited only by the time step). A ramp function, defined by Eq.~\ref{eq:ramp},
is therefore employed to smoothly transition $\Omega^+ (t)$ from zero at $t=0$
to $\Omega^+_0$ at $t=t_0$. In Eq.~\ref{eq:ramp}, `$s$' is a scaling parameter
that determines the steepness of the ramp function.
\begin{equation}
  \Omega^+ (t) = \frac{\Omega^+_0}{2} \left(\frac{\tanh \left(s
  \left(2t/t_0-1\right)\right)}{\tanh (s/t_0)}+1\right)
  \label{eq:ramp}
\end{equation}

\begin{figure}[htb!]
  \incfig[width=0.4 \columnwidth]{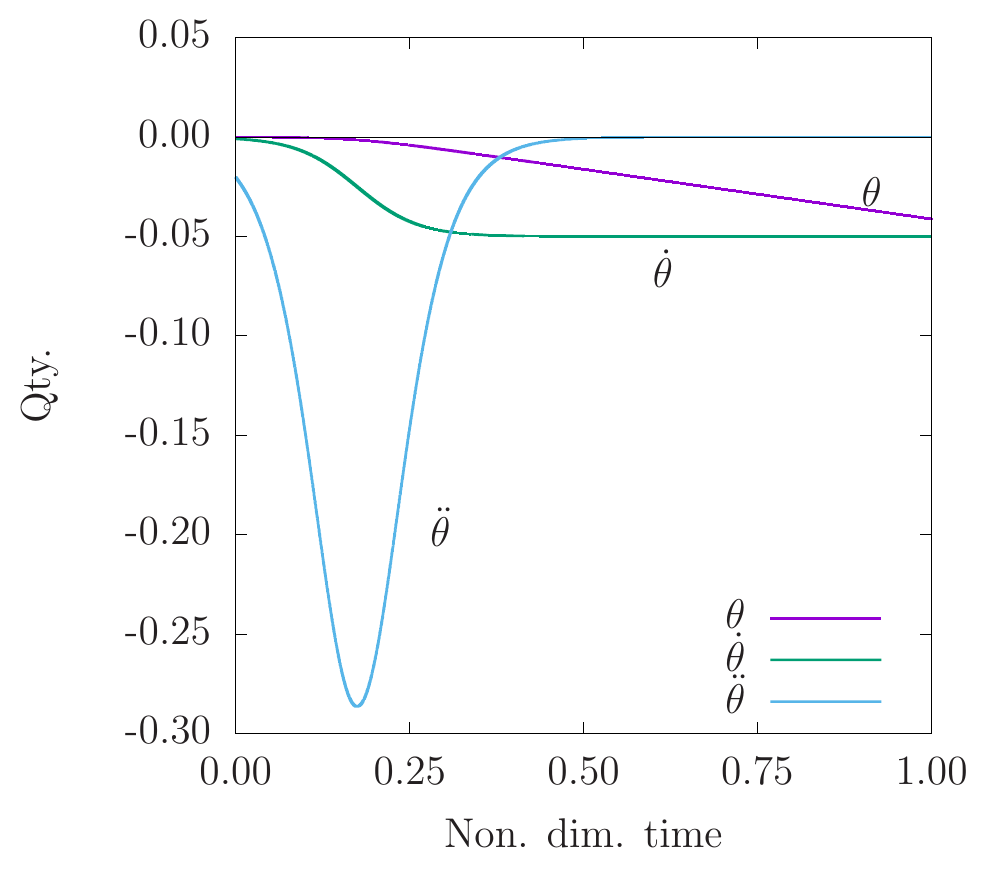}
  \caption{Ramp function used to transition $\dot{\theta} (=\Omega^+)$ from $0$
  to $0.05$, and the associated variations in pitch angle ($\theta$) and
  acceleration ($\ddot{\theta}$); Eq.~\ref{eq:ramp} with $s=2$ and $t_0=1.0$.}
  \label{fig:ramp}
\end{figure}

Figure~\ref{fig:ramp} plots the ramp function (Eq.~\ref{eq:ramp} with $s=2.0$
and $t_0=1.0$) used in the dynamic simulations. The objective is to
transition from $\Omega^+=0$ to the final value $\Omega^+=-0.05$ quickly
without introducing large perturbations due to inertial acceleration.  A
hyperbolic tangent function provides a smooth transition at both end points,
and hence is selected to specify the pitch rate. The transition (ramp) region
is limited by $t_0$ and scaled by $s$; the higher the $s$ value, the quicker
the pitch rate transitions to its final value, but the inertial acceleration is
also high. Since the final pitch rate of $-0.05$ is relatively small, the
effects of inertial acceleration are small and can be ignored.

For $t>t_0$, the airfoil continues to pitch at the constant pitch rate,
$\Omega^+(t)=0.05$, and the angle of attack increases linearly with the pitch
angle, $\theta$. The airfoil goes through various flow stages in the following
sequence during the pitch-up motion: 
\begin{enumerate}
  \item The laminar-to-turbulent boundary layer transition point on the suction
    surface moves upstream towards the leading edge.

  \item A laminar separation bubble (LSB) forms on the suction surface and
    moves closer to the leading edge while simultaneously reducing in size with
    increasing angle of attack. The suction peak ahead of the LSB continues to
    rise; most of the boundary layer on the suction side is turbulent at this
    time.

  \item The LSB ``bursts'' and the suction peak collapses, leading immediately
    to the development of the dynamic stall vortex (DSV).

  \item The DSV convects with the flow. The flow entrainment induced by the DSV
    causes the vorticity in the shear layer in the aft portion of the airfoil to
    roll up into a shear layer vortex (SLV).
  
  \item As the DSV moves downstream, the airfoil pitch-down moment ($-C_M$)
    increases sharply as the lift distribution becomes aft dominant, and moment
    stall occurs.

  \item When the DSV gets close to the trailing edge, the additional lift due to
    the velocity induced by the DSV reduces dramatically, causing lift stall.
\end{enumerate}

\begin{figure}[htb!]
  \incfig[width=\columnwidth]{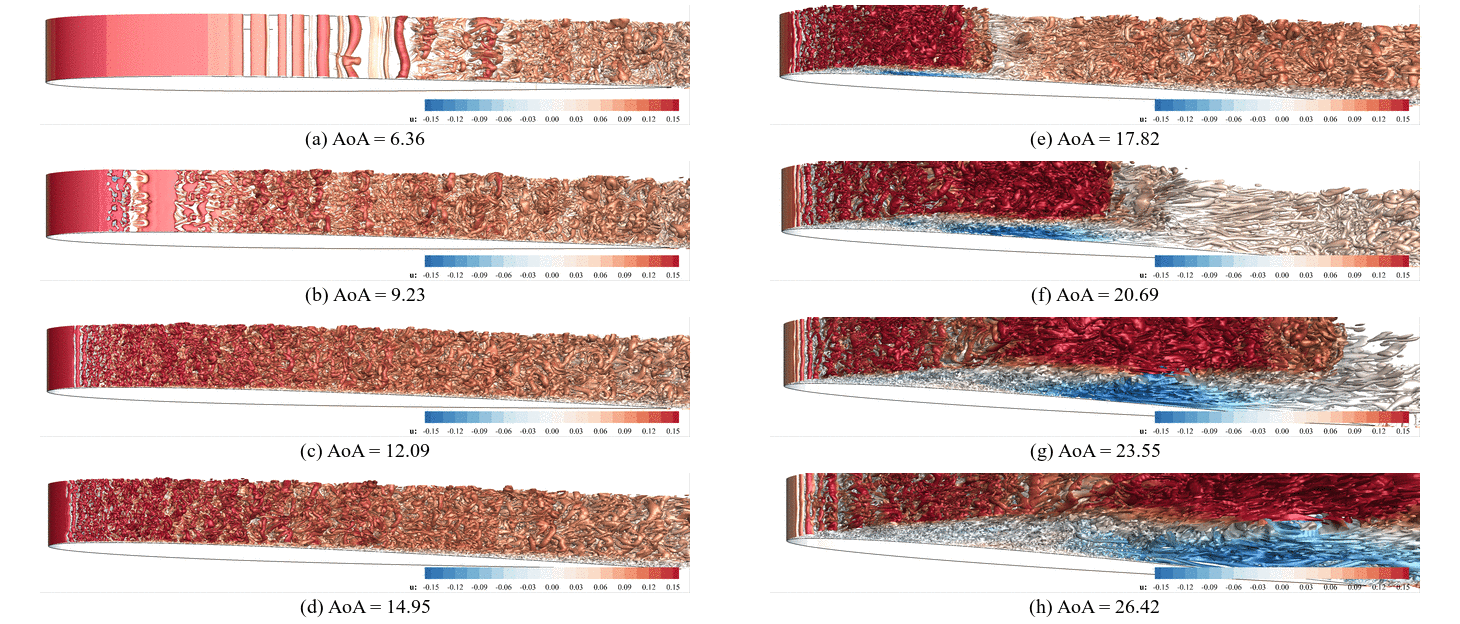}
  \caption{Iso-surfaces of Q-criterion with colored contours of $x-$component
    of flow velocity of the NACA-0012 simulation at various stages of dynamic
    stall.}
  \label{fig:Qcrit_0012}
\end{figure}

This sequence of events can be seen in the snapshots of the FDL3DI predicted
flowfield for the NACA-0012 airfoil in Fig.~\ref{fig:Qcrit_0012}. Each plot in
the figure shows iso-surfaces of the Q-criterion colored by the value of the
$x-$component of flow velocity. The boundary layer transition location can be
clearly seen to have moved upstream in plot (b) compared to plot (a). The LSB
then settles at $x/c \sim 0.06$ and lift continues to increase with $\alpha$.
The LSB bursts somewhere between plots (d) and (e) in Fig.~\ref{fig:Qcrit_0012}
leading to the formation of the DSV, which is seen centered at $x/c \sim 0.2$
in plot (e).  Entrainment of flow by the DSV can be interpreted from the
streamwise elongated eddies seen in plot (f); these are formed because of the
large velocity induced by the DSV impinging on the airfoil and pushing the
residual turbulent boundary layer further downstream, rolling it up into a
shear layer vortex. Plot (f) also marks the beginning of moment stall as the
suction peak moves downstream with the DSV. In plot (h), the DSV is nearing the
trailing edge, marking the onset of lift stall.

%%%%%%%%%%%%%%%%%%%%%%%%%%%%%%%%%%%%%%%%%%%%%%
\subsubsection{Boundary Layer Transition}
\label{sec:transition}
%%%%%%%%%%%%%%%%%%%%%%%%%%%%%%%%%%%%%%%%%%%%%%
%
The transition location is investigated in detail using time accurate pressure
data sampled at several stations along the airfoil suction surface. Pressure
and velocity data is collected at one cell height away from the surface. The
data is collected with a sampling rate of $\Delta f = 25,000\times u/c$, which
is approximately 80,000 data points for each degree of blade rotation.
Aerodynamic pressure coefficient ($C_P$) is averaged along the span to obtain
$\langle C_P \rangle$, which is further low-pass filtered, and the filtered
quantity is denoted by $\langle \widetilde{C_P} \rangle$. Considering $\langle
\widetilde{C_P} \rangle$ as a quantity averaged locally in time, and following
Visbal~\cite{visbal_2014}, we define rms of pressure fluctuations with respect
to this filtered value as ${C_P}_{rms} = \left[ \langle C_P \rangle - \langle
\widetilde{C_P} \rangle \right]^{1/2}$. Early experiments~\cite{lorber_1988}
and some recent measurements at very high sampling rates~\cite{ansell_2017},
have used rms pressure to identify transition location during dynamic stall.
Transition location is identified by a sharp increase in wall pressure
fluctuations.

\begin{figure}[htb!]
  \subfloat[NACA-0009]{\incfig[width=0.47\columnwidth]{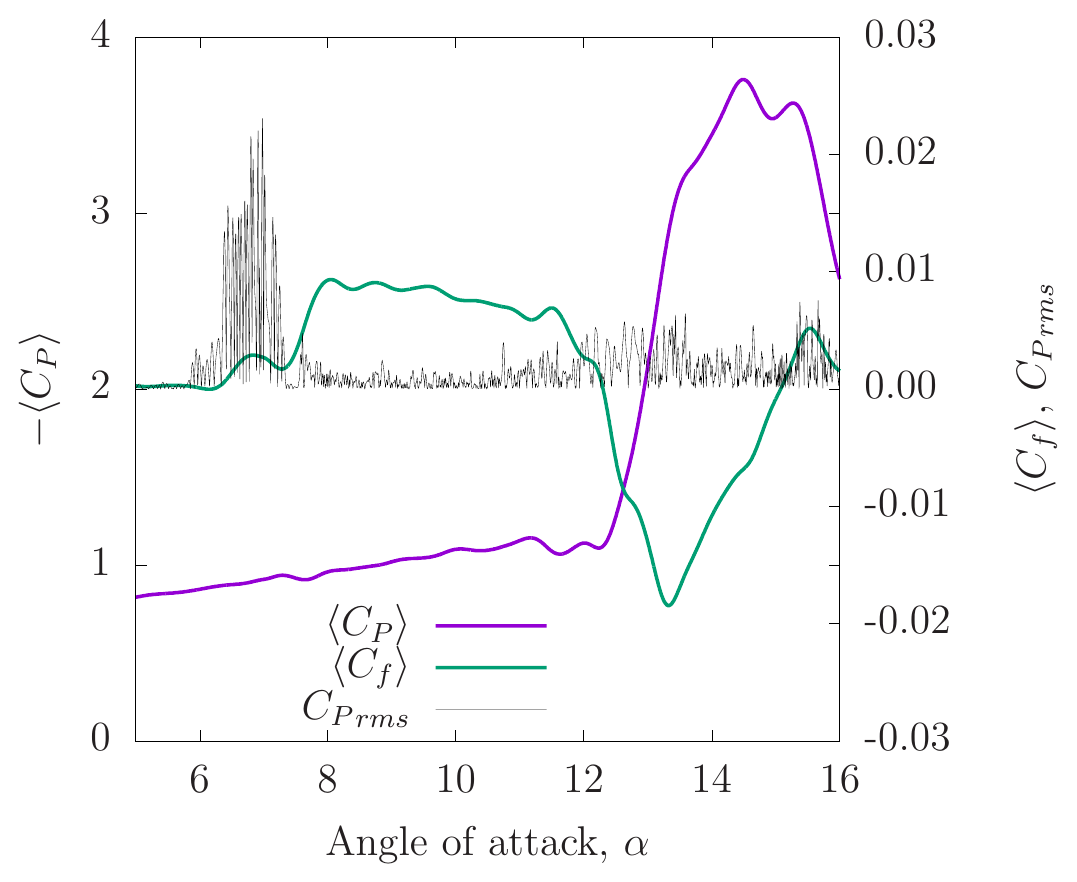}}\;
  \subfloat[NACA-0012]{\incfig[width=0.47\columnwidth]{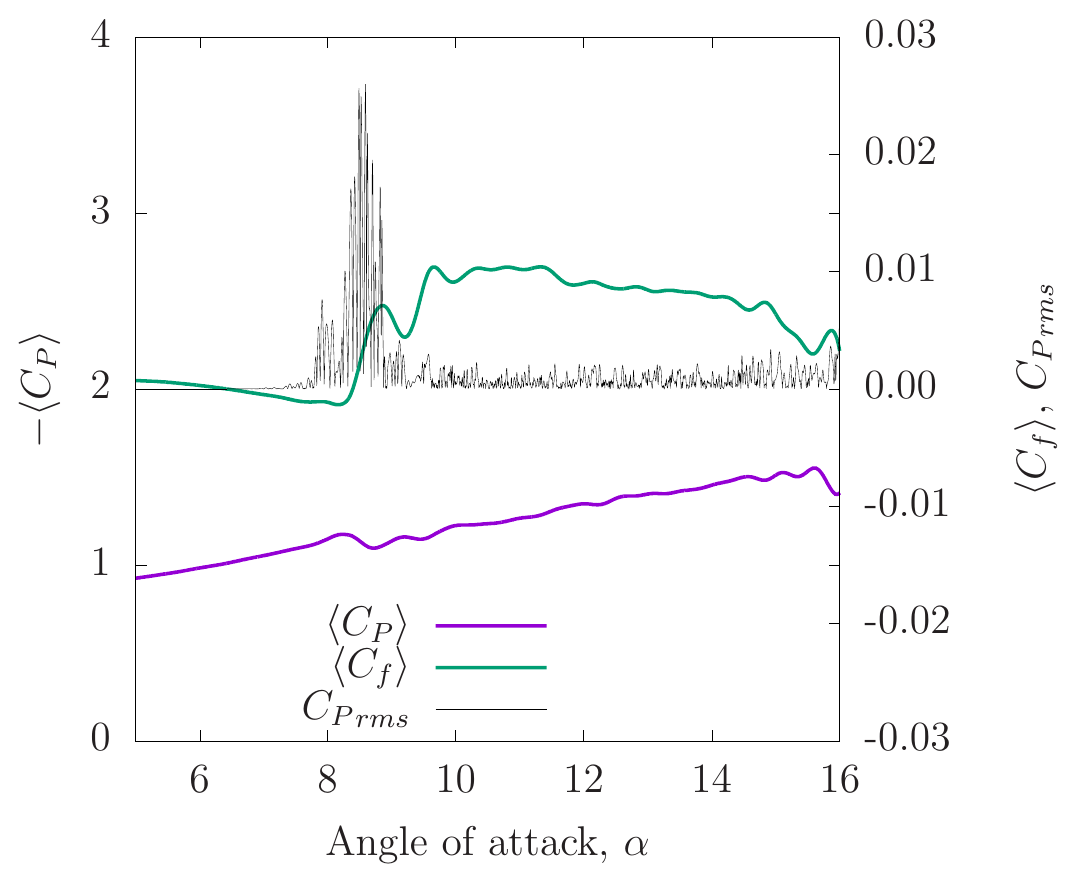}}\\
  \subfloat[NACA-0015]{\incfig[width=0.47\columnwidth]{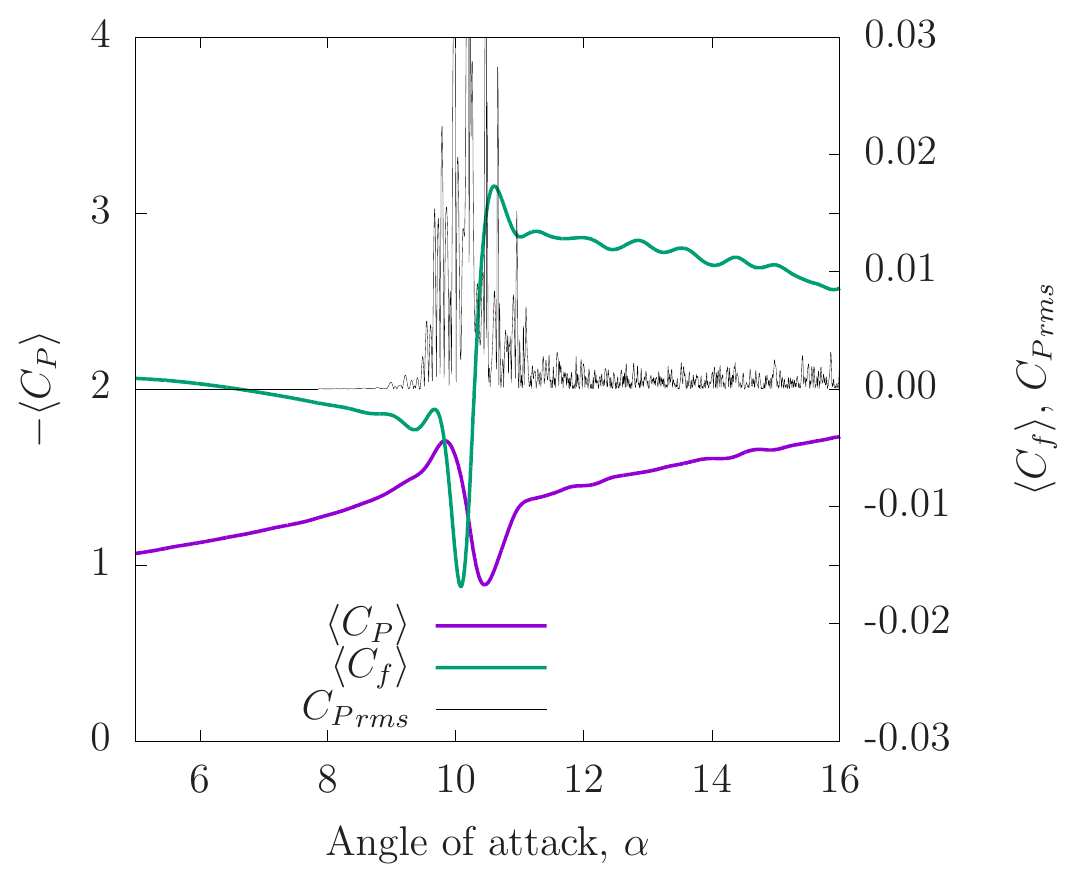}}\;
  \subfloat[NACA-0018]{\incfig[width=0.47\columnwidth]{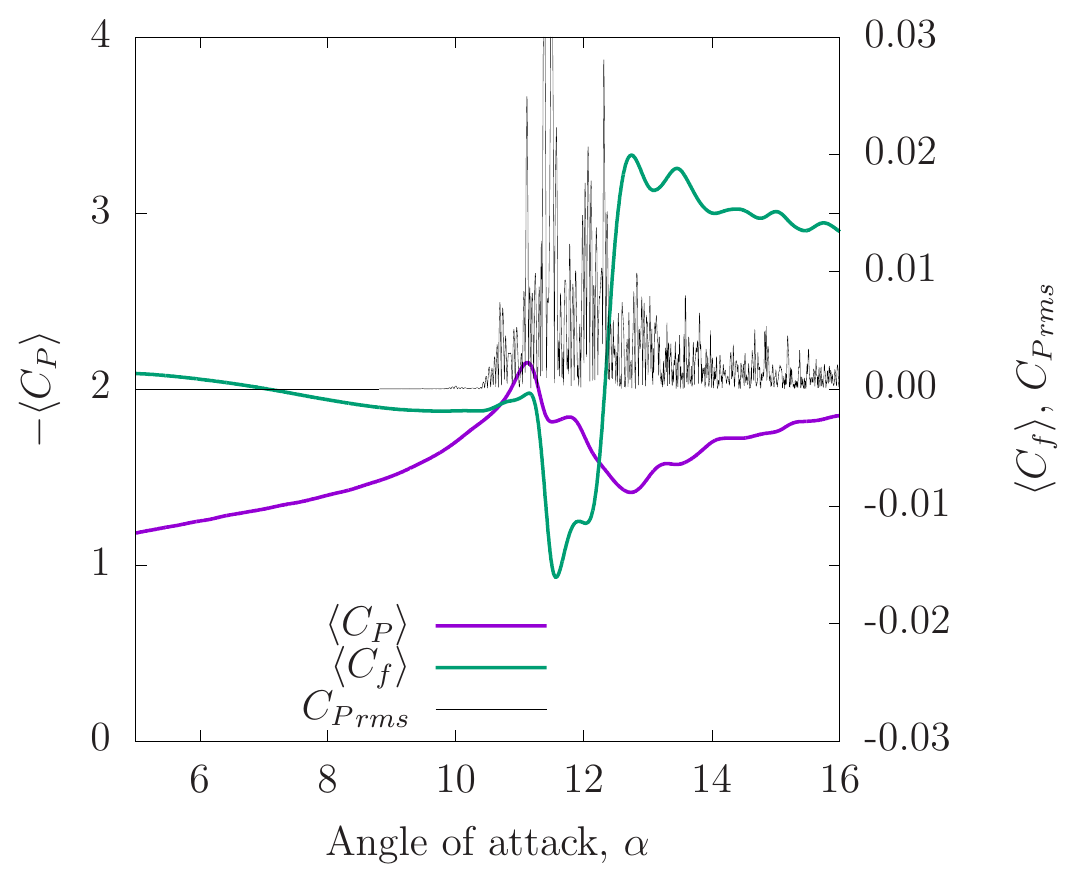}}
  \caption{Identification of transition location using 
    ${C_P}_{rms}$, $\langle \widetilde{C_P} \rangle$, and $\langle C_f \rangle$.} 
  \label{fig:trans_locs}
\end{figure}

Figure~\ref{fig:trans_locs} plots ${C_P}_{rms}$, $\langle \widetilde{C_P}
\rangle$, and $\langle C_f \rangle$ for the four airfoils at $x/c=0.02$ as they
go through the pitch-up maneuver. A large increase in ${C_P}_{rms}$ (defined
w.r.t. $\langle \widetilde{C_P} \rangle$ is clearly visible for each airfoil,
which coincides with the angle of attack where $\langle C_f \rangle$ increases
sharply. For the simulations considered, the $\langle C_f \rangle$ dips
negative before the transition location, which is due to the reverse flow
inside the LSB. The sharp jumps observed in ${C_P}_{rms}$ and $\langle C_f
\rangle$ are consistent with the increase in fluctuations due to the boundary
layer turning turbulent. At the transition location, a dip in suction pressure
($\langle \widetilde{C_P} \rangle$) is also observed, consistent with the
measurements reported in Ref.~\cite{ansell_2017}.

%%%%%%%%%%%%%%%%%%%%%%%%%%%%%%%%%%%%%%%%%%%%%%%%%%%%
\subsubsection{Lift, Drag, and Moment Variations}
\label{sec:clcdcm}
%%%%%%%%%%%%%%%%%%%%%%%%%%%%%%%%%%%%%%%%%%%%%%%%%%%%
%
The four airfoils tested here more-or-less follow the same general pattern as
the pitch angle is increased through stall, although there are considerable
differences in the unsteady lift increase, local pressure peaks, and the amount
of trailing edge separation before stall occurs. These differences are
discussed next.

\begin{figure}[htb!]
  \centering
  \subfloat[$c_l$]{\incfig[width=0.32\columnwidth]{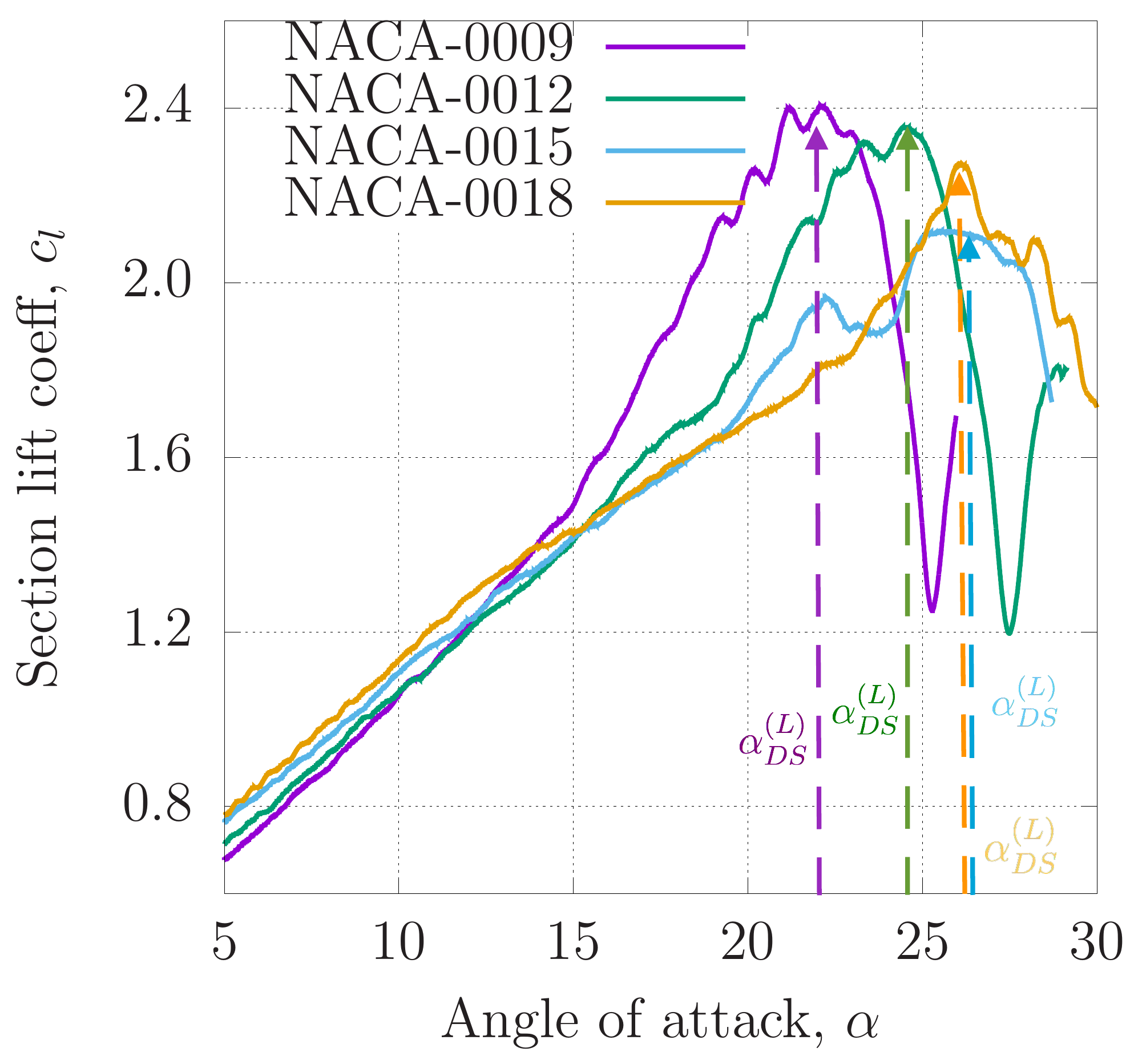}}\;
  \subfloat[$c_d$]{\incfig[width=0.32\columnwidth]{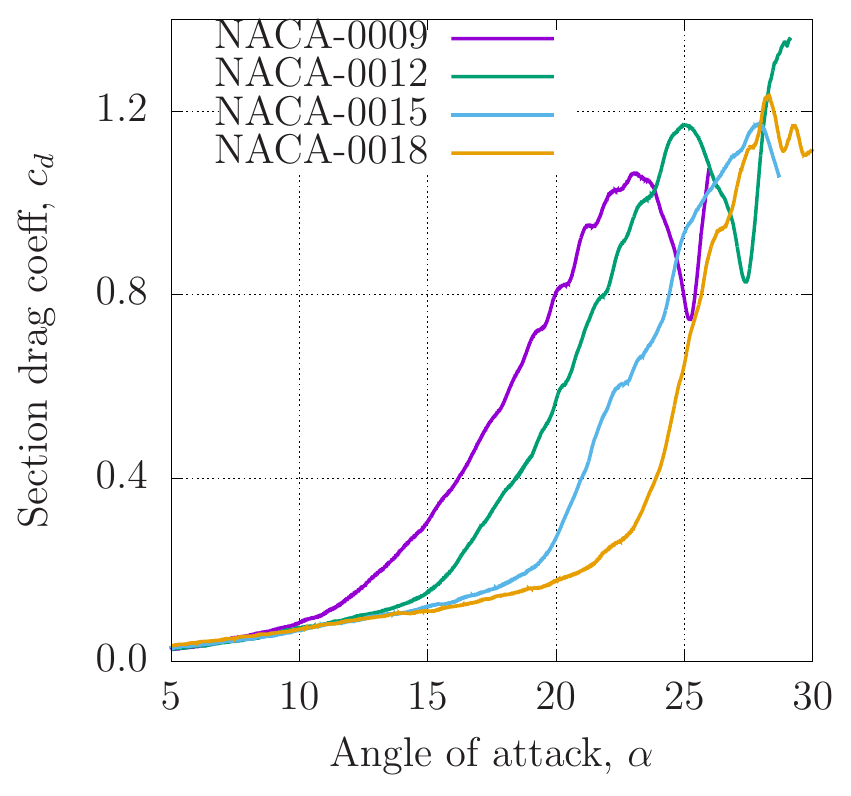}}\;
  \subfloat[$c_m$]{\incfig[width=0.32\columnwidth]{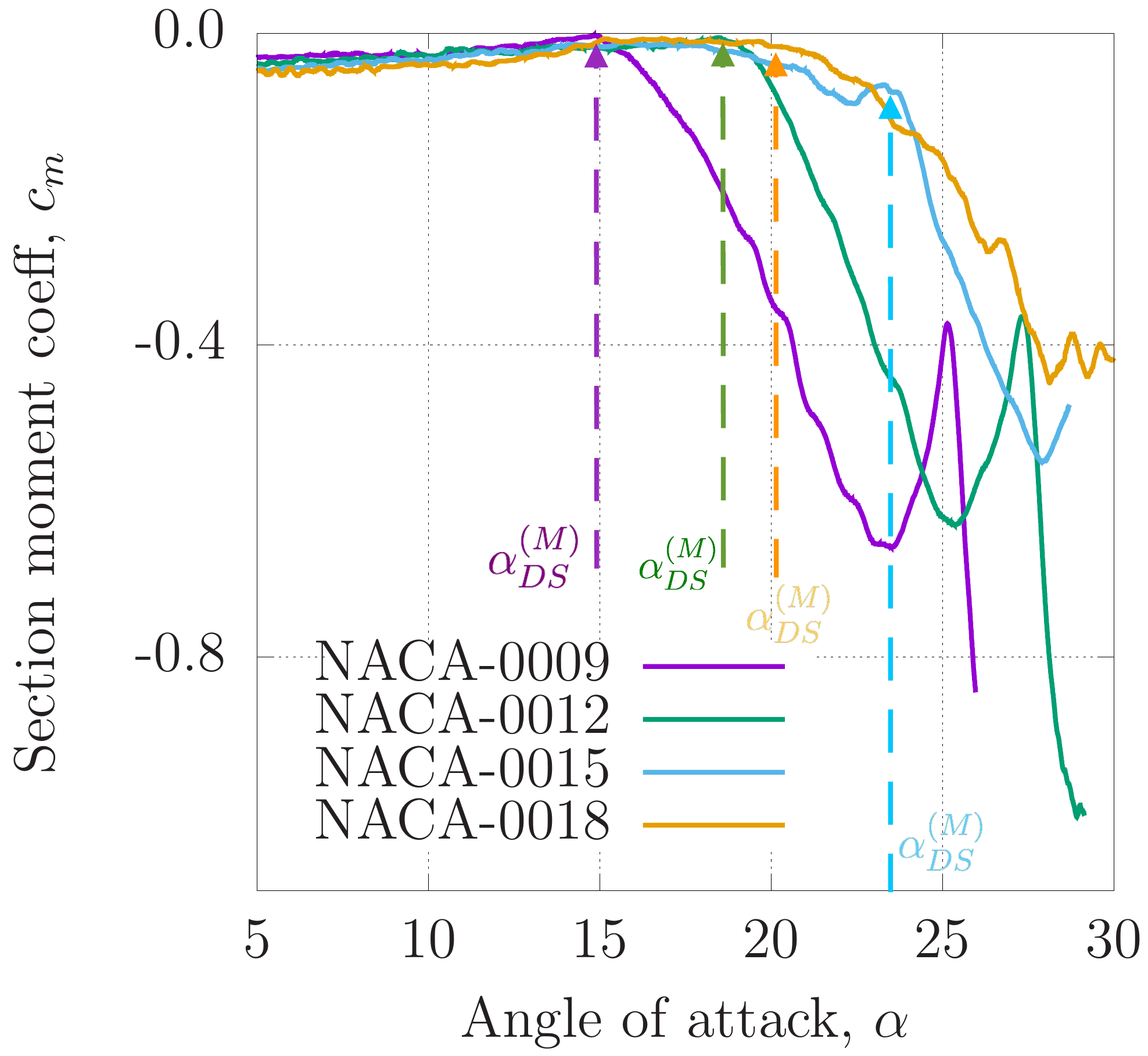}}
  \caption{Sectional lift-, drag-, and moment coefficients as functions of angle of
  attack during a constant pitch-rate maneuver.}
  \label{fig:dyn_ClCdCm}
\end{figure}

Figure~\ref{fig:dyn_ClCdCm} compares the dynamic section lift-, drag-, and
moment coefficients for the four simulated airfoils as they undergo the
constant-rate pitching motion. We focus first on the NACA-0012 simulation. The
slope of the $c_l-\alpha$ curve increases around $\alpha=18^\circ$, which is due to
the strengthening of the DSV and the associated increase in lift. This is
immediately followed by moment stall, marked by the strong divergence in the
$c_m-\alpha$ curve. As explained earlier, the sharp increase in pitch-down
moment is due to the progressive aft propagation of loading induced by the DSV.
At around $\alpha=25^\circ$ the DSV has propagated close to the trailing edge and
away from the airfoil. As a result the lift induced by the DSV reduces
dramatically and lift stall occurs.

Comparing the sectional lift, drag and moment for the four airfoils (see
Fig.~\ref{fig:dyn_ClCdCm} and Table~\ref{tab:alpha_SSvsDS}) shows that the
largest increase in lift and pitch-down moment due to airfoil motion (dynamic
stall), is observed for the NACA-0009 airfoil; the smallest increase in lift is
observed for the NACA-0015 airfoil; while the NACA-0018 experiences the
smallest increase in pitch-down moment. The increase in unsteady lift is
measured as the difference of $c_{l,max}$ between dynamic- and static stall.
The values for dynamic stall are obtained using FDL3DI while the corresponding
static values are obtained using XFOIL. The angle of attack beyond which the
drag coefficient increases rapidly, increases monotonically with airfoil
thickness -- the thinnest airfoil showing the divergence at much smaller
$\alpha$ than the thick airfoils. While unsteady loads reduce with increasing
airfoil thickness, stall delay (as measured by the difference in $\alpha$ where
dynamic stall occurs versus where static stall occurs) remains nearly
unchanged. The static stall $\alpha$ values for the four airfoils are also
obtained using XFOIL.

\begin{table}[htb!]
\centering
\caption{Angle of attack values at which static stall and dynamic stall occurs
  (denoted by $\alpha_{SS}$ and $\alpha_{DS}$ respectively) for different
  airfoils.  Moment stall and lift stall values are indicated separately.
  Static stall values are obtained using XFOIL whereas dynamic stall values are
  from FDL3DI simulations.}
\label{tab:alpha_SSvsDS}
\begin{tabular}{r|c|l|l|l}
\multirow{2}{*}{} & \textbf{Moment Stall} & \multicolumn{3}{c}{\textbf{Lift Stall}} \\ \cline{2-5}  \cline{2-5}
                  &                   $\alpha^{(M)}_{DS}$ & $\alpha^{(L)}_{DS}$ &                   $\alpha^{(L)}_{SS}$ & $\Delta \alpha^{(L)} = \alpha^{(L)}_{DS} - \alpha^{(L)}_{SS}$  \\ \hline  \hline
        NACA 0009 & \cellcolor[HTML]{EFEFEF}15.0           &   22.2               & \cellcolor[HTML]{EFEFEF}  10.7         &    11.5                                        \\ \hline
        NACA 0012 & \cellcolor[HTML]{EFEFEF}18.7           &   24.6               & \cellcolor[HTML]{EFEFEF}  13.7         &    10.9                                        \\ \hline
        NACA 0015 & \cellcolor[HTML]{EFEFEF}23.5           &   25.5               & \cellcolor[HTML]{EFEFEF}  15.0         &    10.5                                        \\ \hline
        NACA 0018 & \cellcolor[HTML]{EFEFEF}22.0           &   26.0               & \cellcolor[HTML]{EFEFEF}  17.0         &     9.0                                        \\ \hline \hline
\end{tabular}
\end{table}

%%%%%%%%%%%%%%%%%%%%%%%%%
\subsubsection{Effect of Finite Span in Simulations}
\label{sec:coherence}
%%%%%%%%%%%%%%%%%%%%%%%%%
%
The span of the simulated airfoil geometries is equal to 10 percent of the
airfoil chord length. Periodic boundary conditions are employed in the span
direction. The impact of using finite span length is assessed by investigating
spanwise coherence at different stages during the pitch-up maneuver. Magnitude
squared coherence $\gamma^2(\Delta z)$ is defined as
\begin{align}
  \gamma^2(\Delta z) &= \frac{\langle \abs{S_{xy}}^2 \rangle}{\langle S_{xx}\rangle \langle S_{yy} \rangle}
  \label{eq:coherence}
\end{align}
where $S_{xy}=\int_{-\infty}^{\infty} \exp(-i \omega \tau) R_{xy}(\tau)\,{\rm
d} \tau$ is the cross-spectral density of pressures between two points along
the span separated by $\Delta z$, at a fixed chord-wise location of $x/c=0.5$;
$S_{xx} = \int_{-\infty}^{\infty} \exp(-i \omega \tau) R_{xx}(\tau)\,{\rm d}
\tau$ and $S_{yy} = \int_{-\infty}^{\infty} \exp(-i \omega \tau) R_{yy}(\tau)
\,{\rm d} \tau$ are power spectral densities at each of the two points.  The
cross-spectral and power spectral densities are respectively the Fourier
transforms of the cross-correlation ($R_{xy}(\tau)$) and auto-correlation
($R_{xx}(\tau)$) functions of the signals (pressure time history). The angular
brackets in Eq.~\ref{eq:coherence} denote ensemble average, which is reduced to
time averaging here by assuming ergodicity.

\begin{figure}[htb!]
  \subfloat[Interval 1]{\incfig[width=0.9\columnwidth]{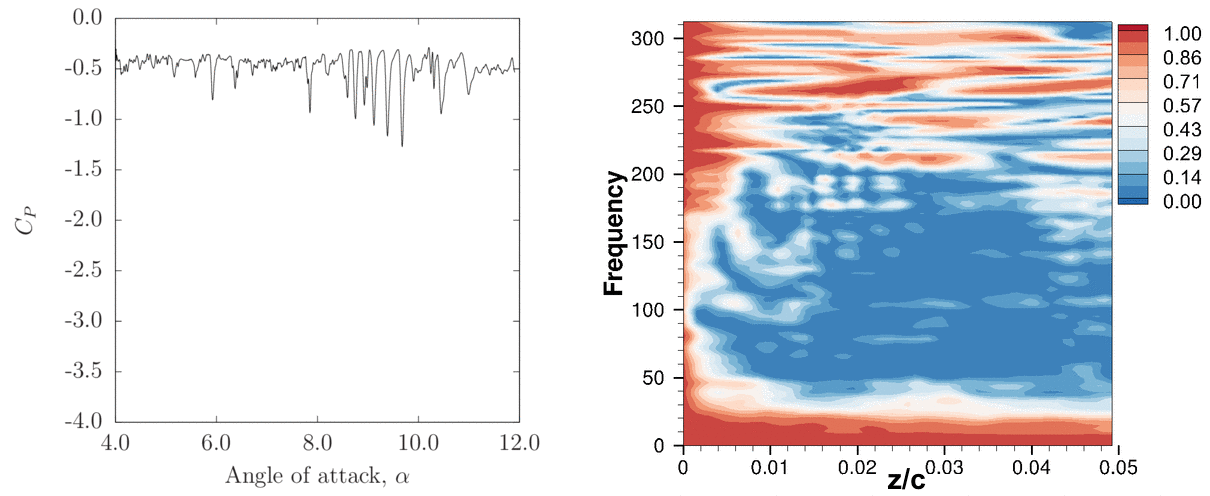}} \\
  \subfloat[Interval 2]{\incfig[width=0.9\columnwidth]{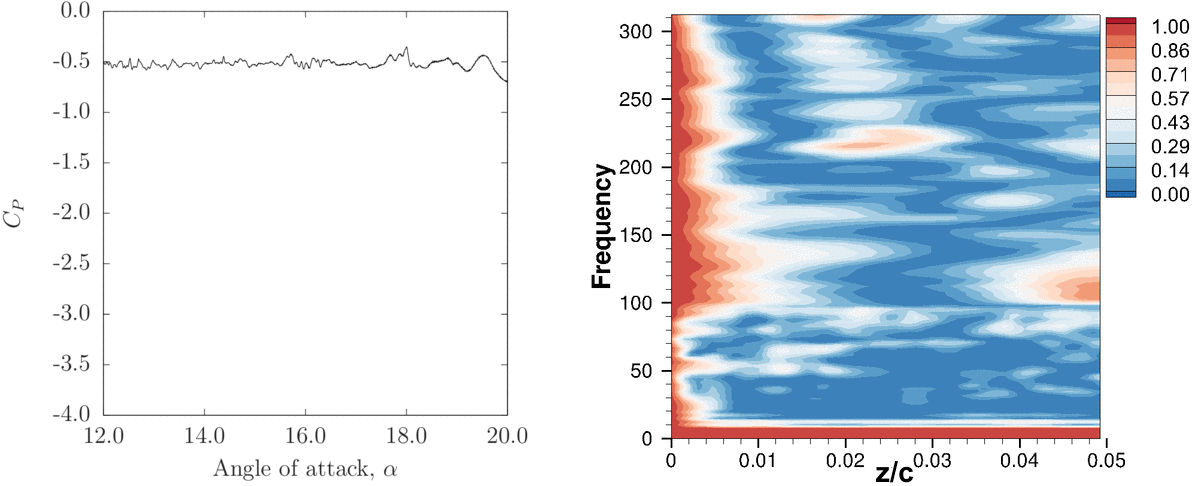}} \\
  \subfloat[Interval 3]{\incfig[width=0.9\columnwidth]{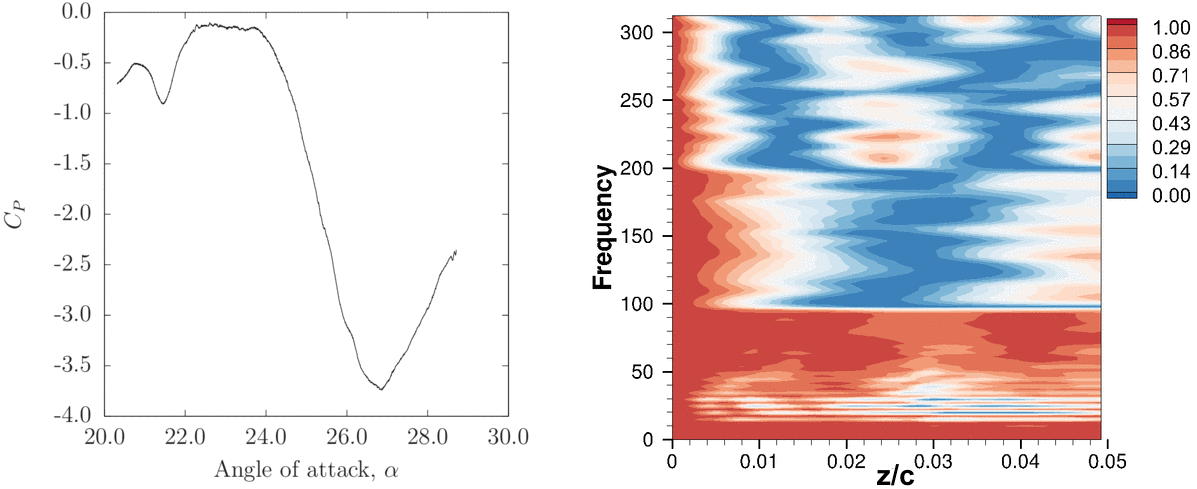}}
  \caption{Spanwise coherence of $C_P$ at $x/c=0.5$ (shown for the NACA 0015
    airfoil) during pitch-up maneuver. The entire maneuver is divided into three
    intervals to assess adequacy of the span length in the simulations.}
  \label{fig:coherence}
\end{figure}

The entire pitch-up maneuver is divided into three time intervals. The left
plots in Fig.~\ref{fig:coherence} plot the pressure signal in the time domain
at a reference point on the airfoil suction surface ($x/c=0.5; z/c=0$) for
these three intervals. Magnitude square coherence, $\gamma^2(\Delta z)$ plots
for each of these intervals are shown on the right in Fig.~\ref{fig:coherence}.
The first interval is characterized by strong instability modes that ultimately
cause boundary layer transition on the suction surface. These instability modes
are highly correlated in the span direction; they are essentially
two-dimensional. The coherence plot for this time interval shows high spanwise
correlation at several high frequencies corresponding to these essentially 2-D
modes.

In the second interval, the boundary layer is turbulent at the selected
chord-wise location, and dynamic stall onset occurs towards the very end
of the interval ($\alpha \sim 20^\circ$). The corresponding coherence plot shows
relatively small spanwise coherence, suggesting that the simulated span length
is sufficient to investigate onset of dynamic stall.

In the third time interval, the DSV convects over the chordwise location,
$x/c=0.5$ and the airfoil experiences deep stall. Very large coherence is
observed at low frequencies corresponding to the large-scale, slow-moving DSV.
The span length of 10\% chord is therefore not sufficient to study post-stall
airfoil behavior, however simulation of interval 2, where stall onset occurs,
can be carried out on this finite-span geometry model. This conclusion is
corroborated by the results for simulations performed for different span
lengths.~\cite{visbal_2017}

%%%%%%%%%%%%%%%%%%%%%%%%%
\subsubsection{Onset of Dynamic Stall}
\label{sec:stallMechanisms}
%%%%%%%%%%%%%%%%%%%%%%%%%
%
Based on the nature of the boundary layer separation leading to dynamic stall,
McCroskey~\etal~\cite{mccroskey1981} classified dynamic stall (see
Fig.~4 in Ref.~\cite{mccroskey1981}) into the following categories:
\begin{enumerate}

  \item {\em Leading edge stall} can occur in one of two ways - (a) the LSB may
    ``burst'' as the adverse pressure gradient becomes too high and the
    separated shear layer fails to re-attach, leading to formation of the DSV,
    or (b) via an abrupt forward propagation of flow reversal to the leading
    edge.

  \item {\em Trailing edge stall} initiates with flow reversal near the
    trailing edge. The reverse flow region gradually expands as the separation
    location moves upstream with increasing angle of attack. Once the
    separation point reaches close to the leading edge, the reverse flow region
    covers most of the airfoil. The DSV then forms at the leading edge and
    convects downstream and away from the airfoil.

  \item {\em Thin airfoil stall} is said to occur when the LSB progressively
    lengthens and covers the entire airfoil.

  \item {\em Mixed stall} can occur in two ways: (a) flow separation occurs
    simultaneously near the leading and trailing edges and the separation
    points move toward each other and merge near mid-chord, or (b) flow
    separation occurs near mid-chord, the separation point subsequently
    bifurcates with one branch moving upstream and the other downstream.
\end{enumerate}
\clearpage

\begin{figure}[htb!]
  \subfloat[NACA-0009]{\incfig[width=0.45\columnwidth]{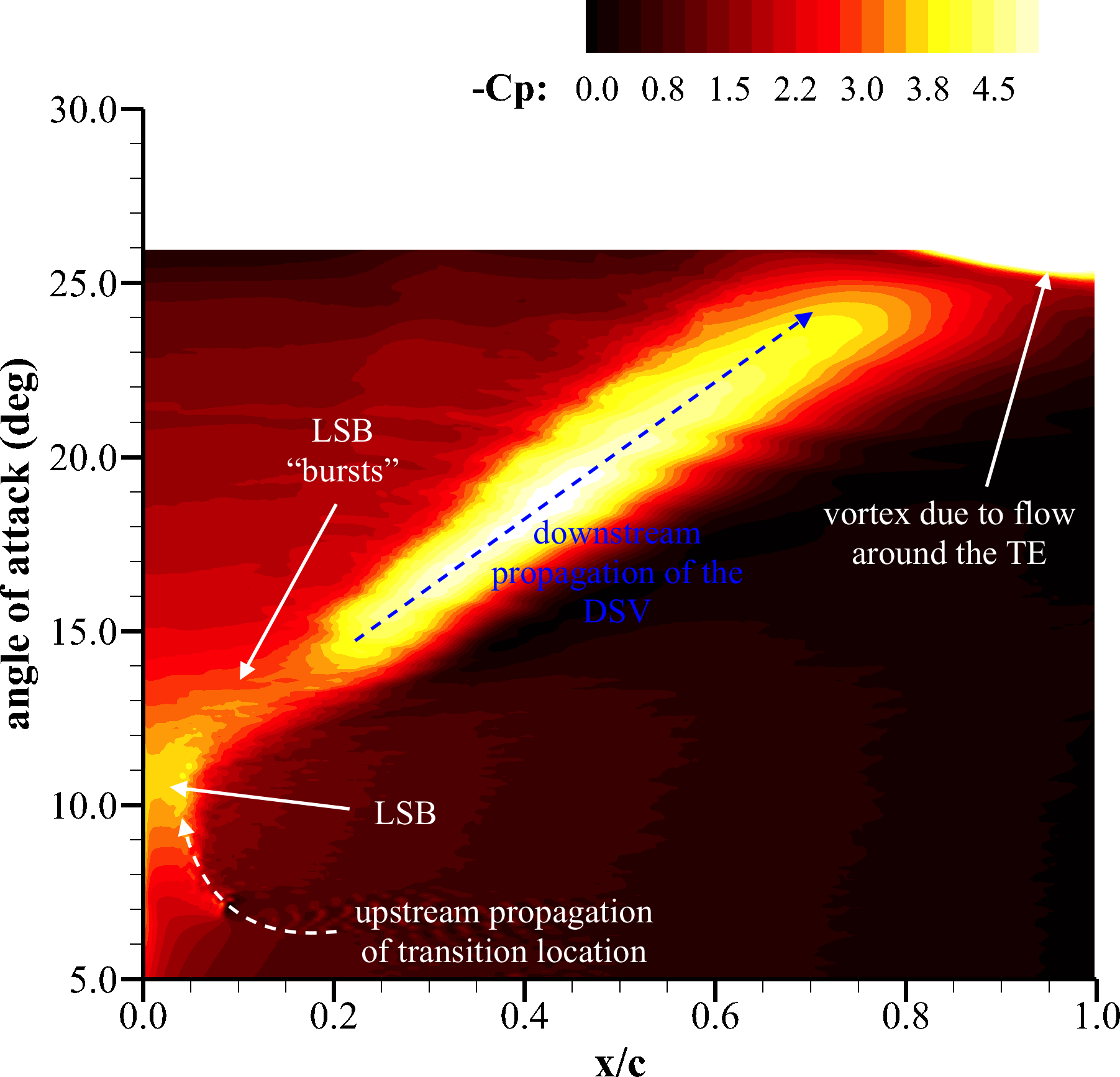}} \qquad
  \subfloat[NACA-0012]{\incfig[width=0.45\columnwidth]{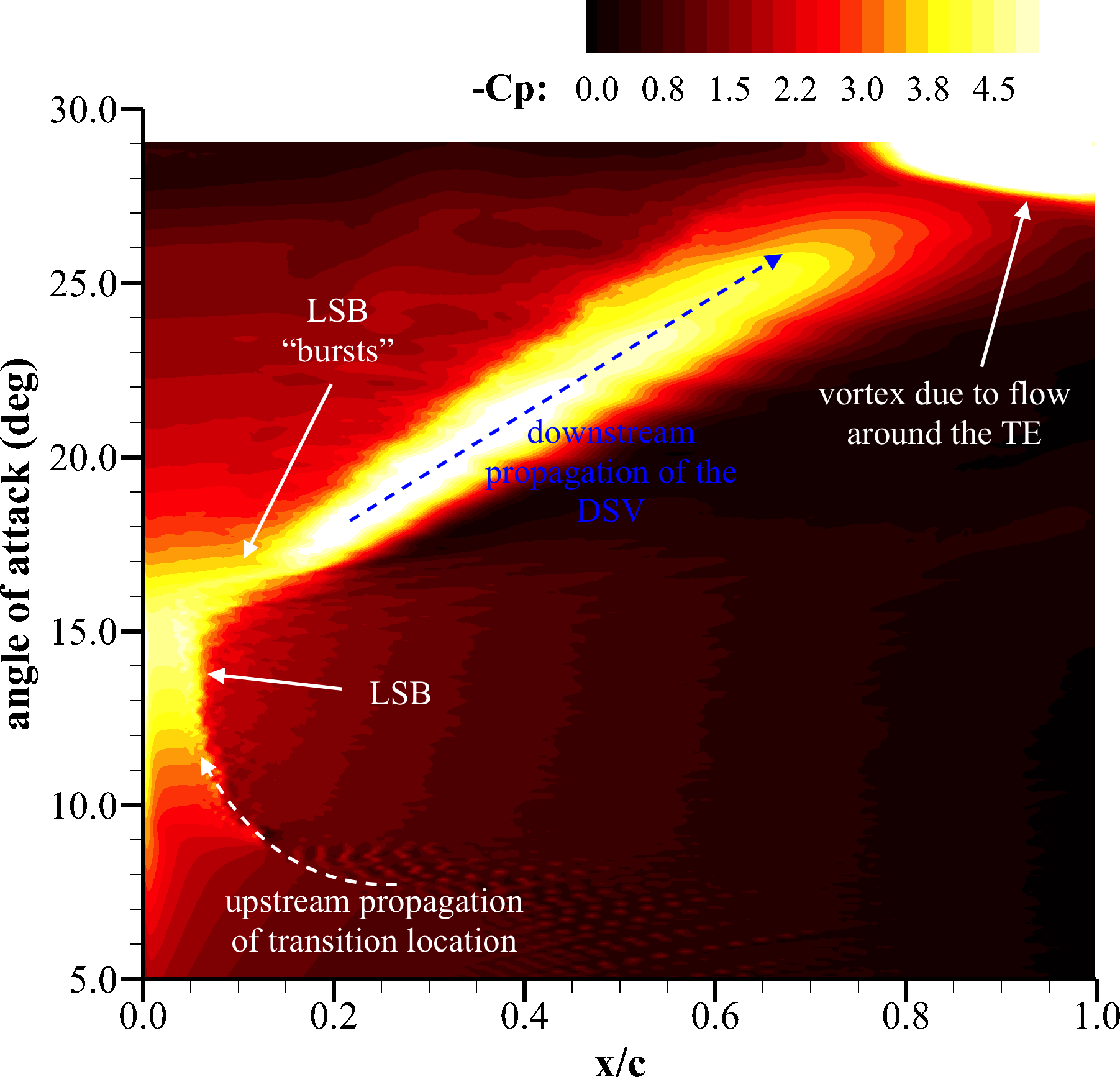}} \\
  \subfloat[NACA-0015]{\incfig[width=0.45\columnwidth]{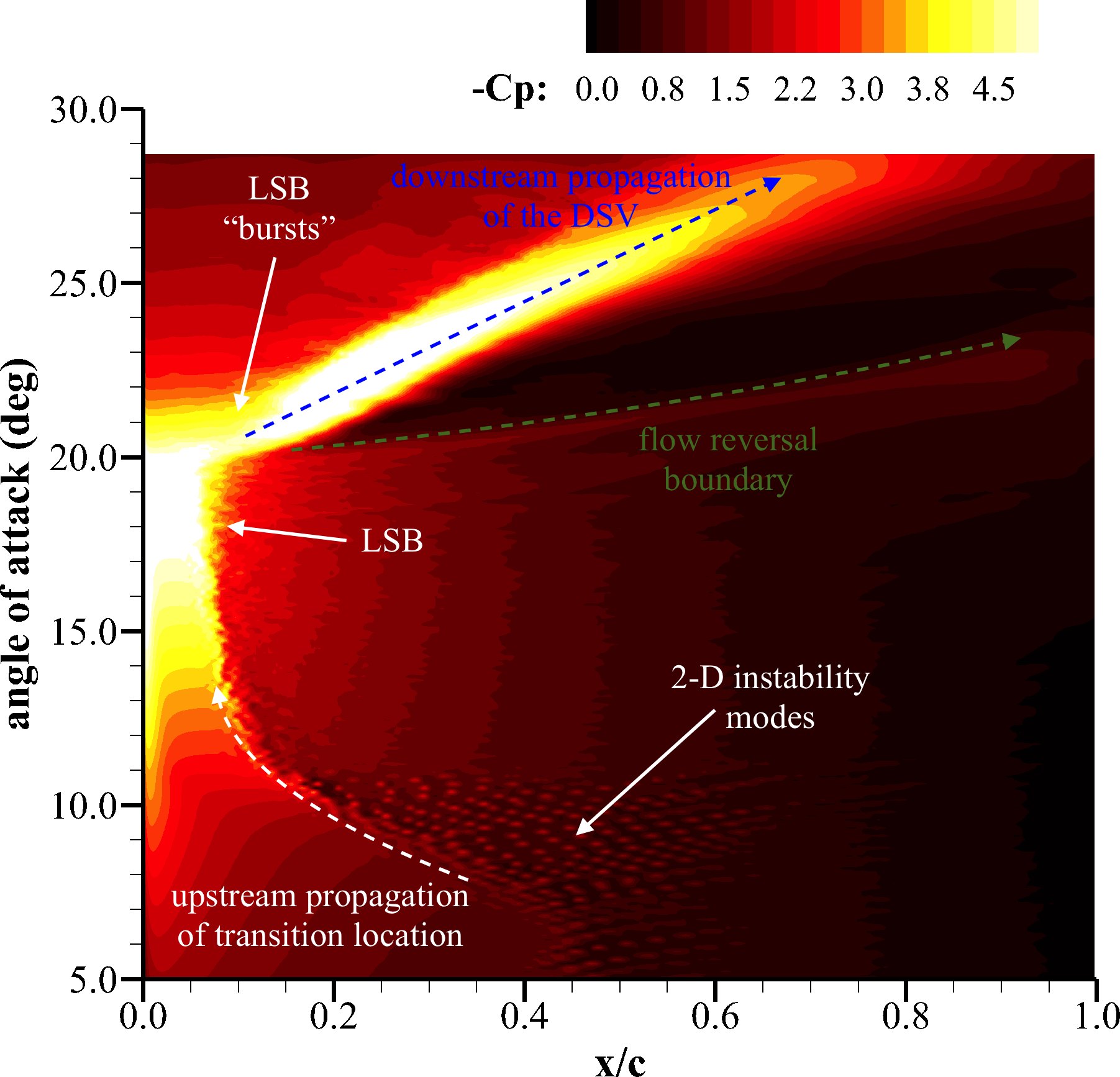}} \qquad
  \subfloat[NACA-0018]{\incfig[width=0.45\columnwidth]{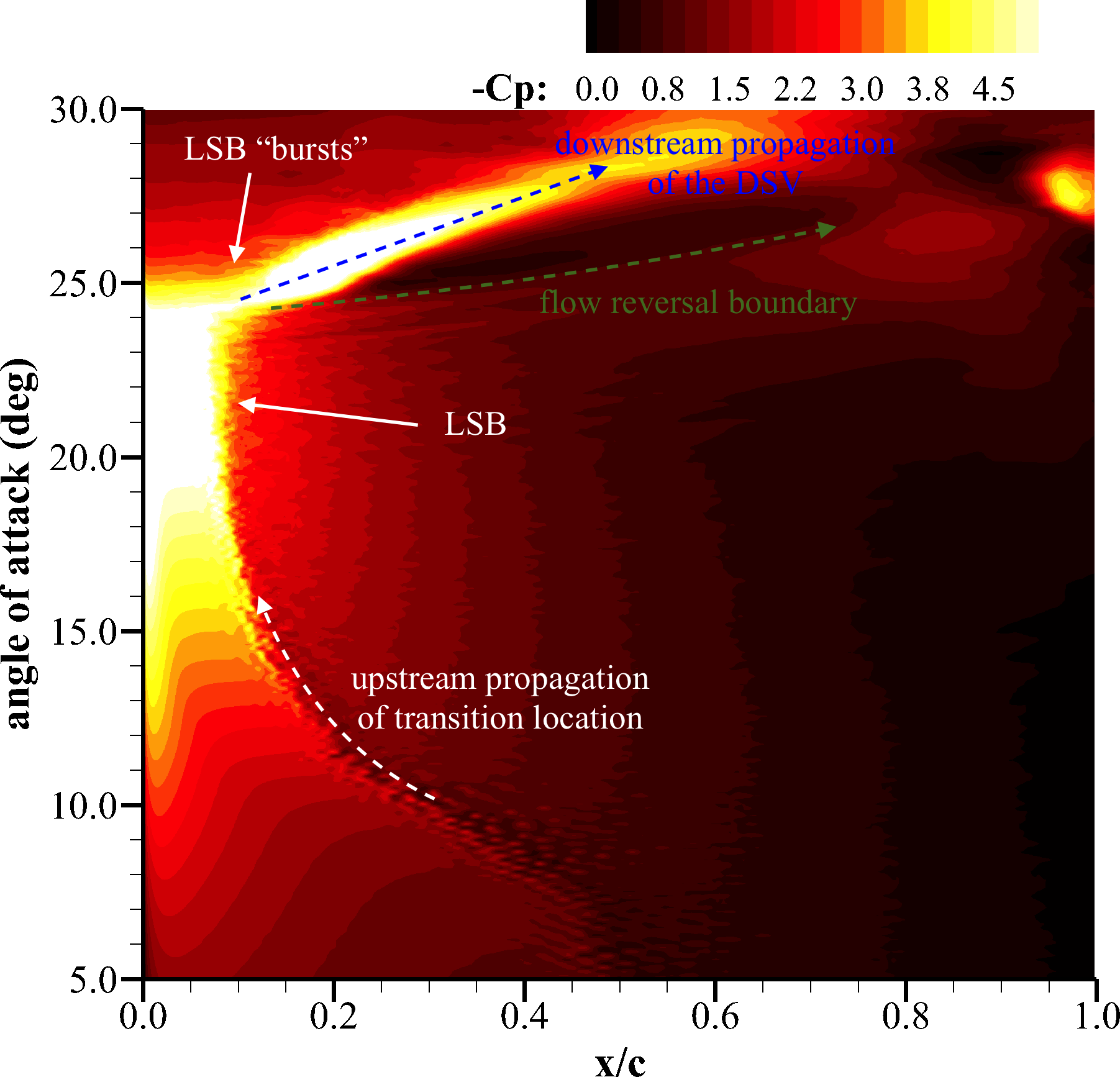}} \;
  \caption{Contours of span-averaged pressure coefficient ($\langle C_P
    \rangle$) on the suction side of the four airfoils through the constant-rate
    pitch-up motion.}
  \label{fig:Cp_contours}
\end{figure}

\begin{figure}[htb!]
  \subfloat[NACA-0009]{\incfig[width=0.45\columnwidth]{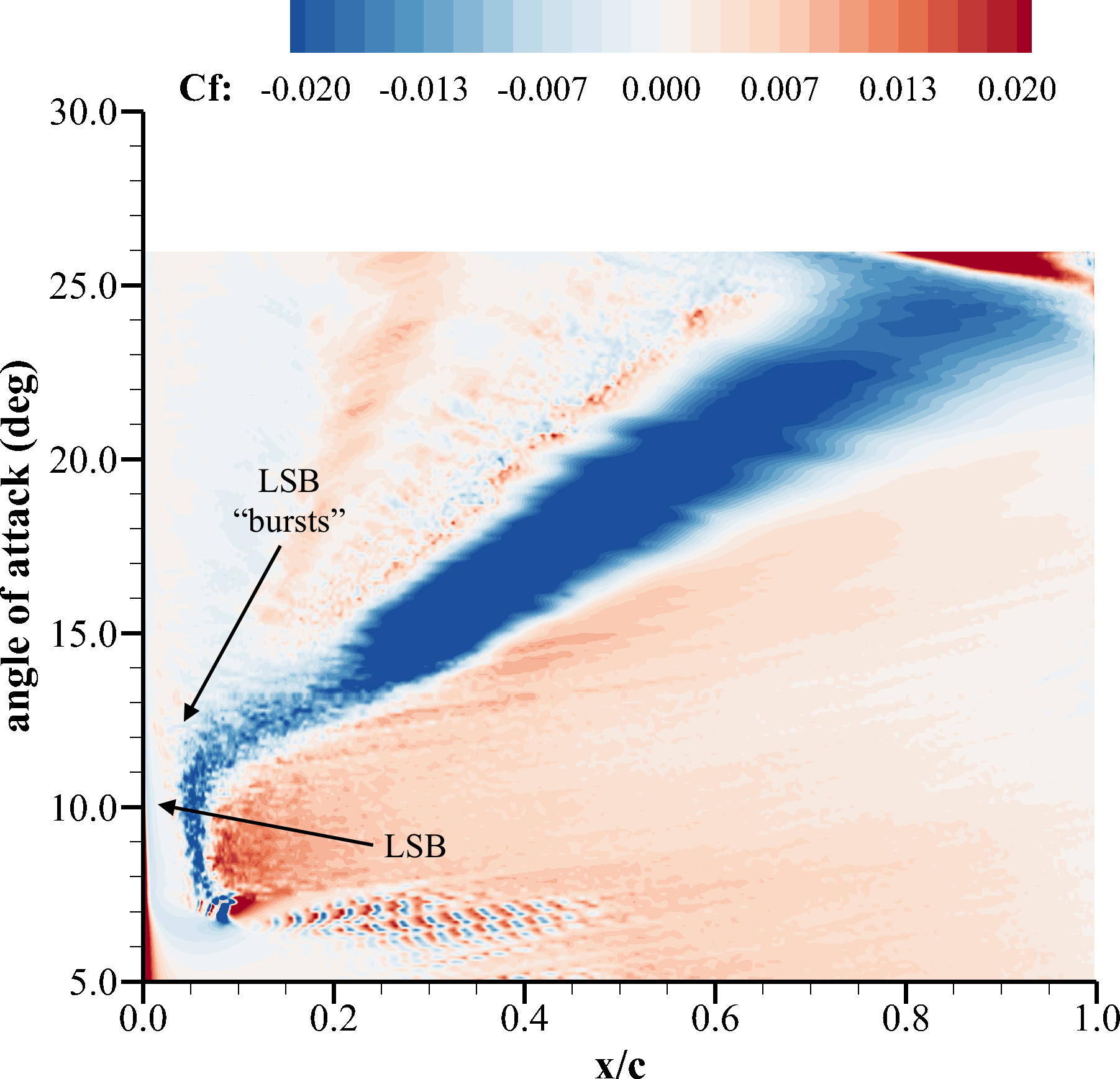}} \qquad
  \subfloat[NACA-0012]{\incfig[width=0.45\columnwidth]{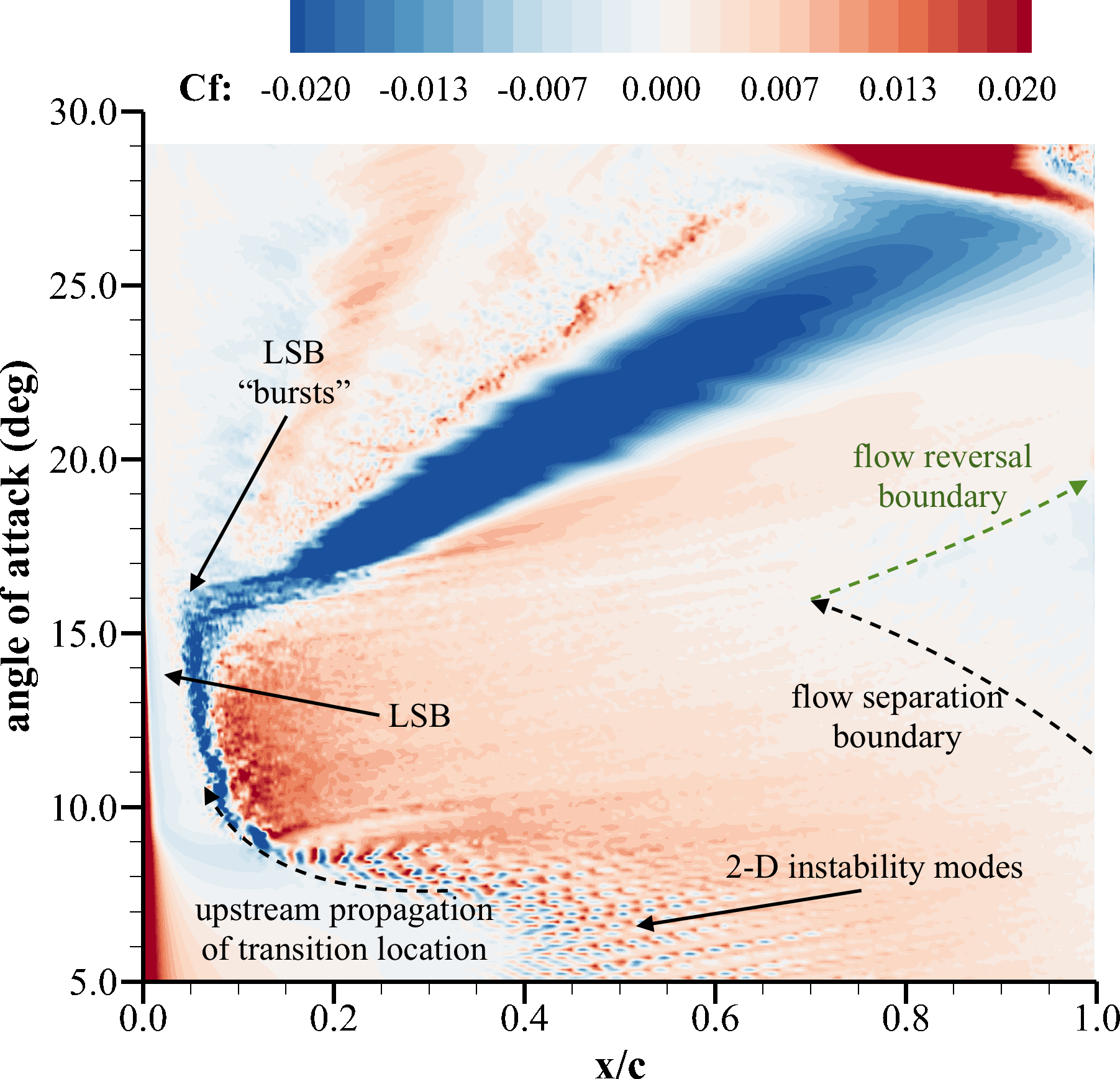}} \\
  \subfloat[NACA-0015]{\incfig[width=0.45\columnwidth]{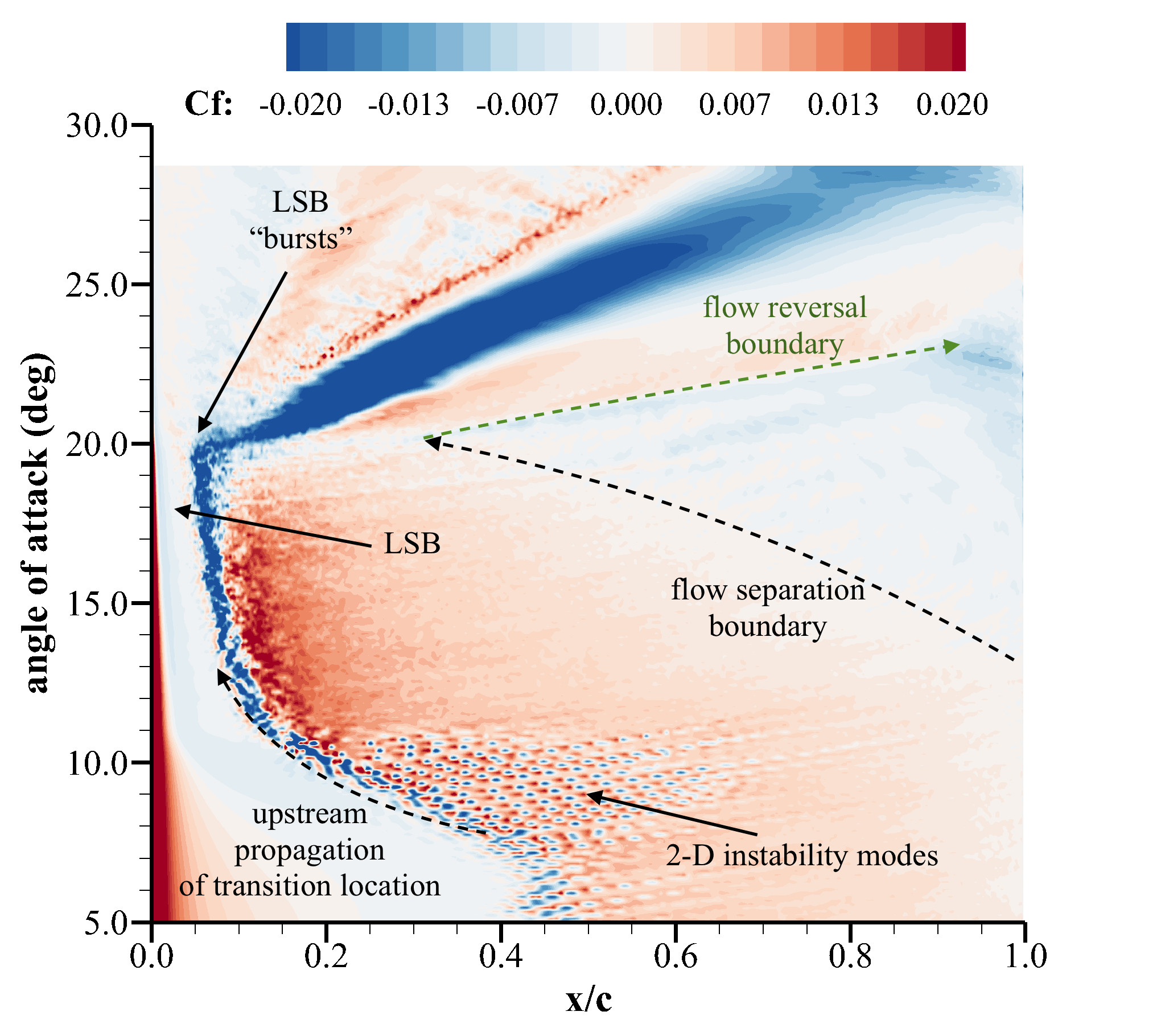}} \qquad
  \subfloat[NACA-0018]{\incfig[width=0.45\columnwidth]{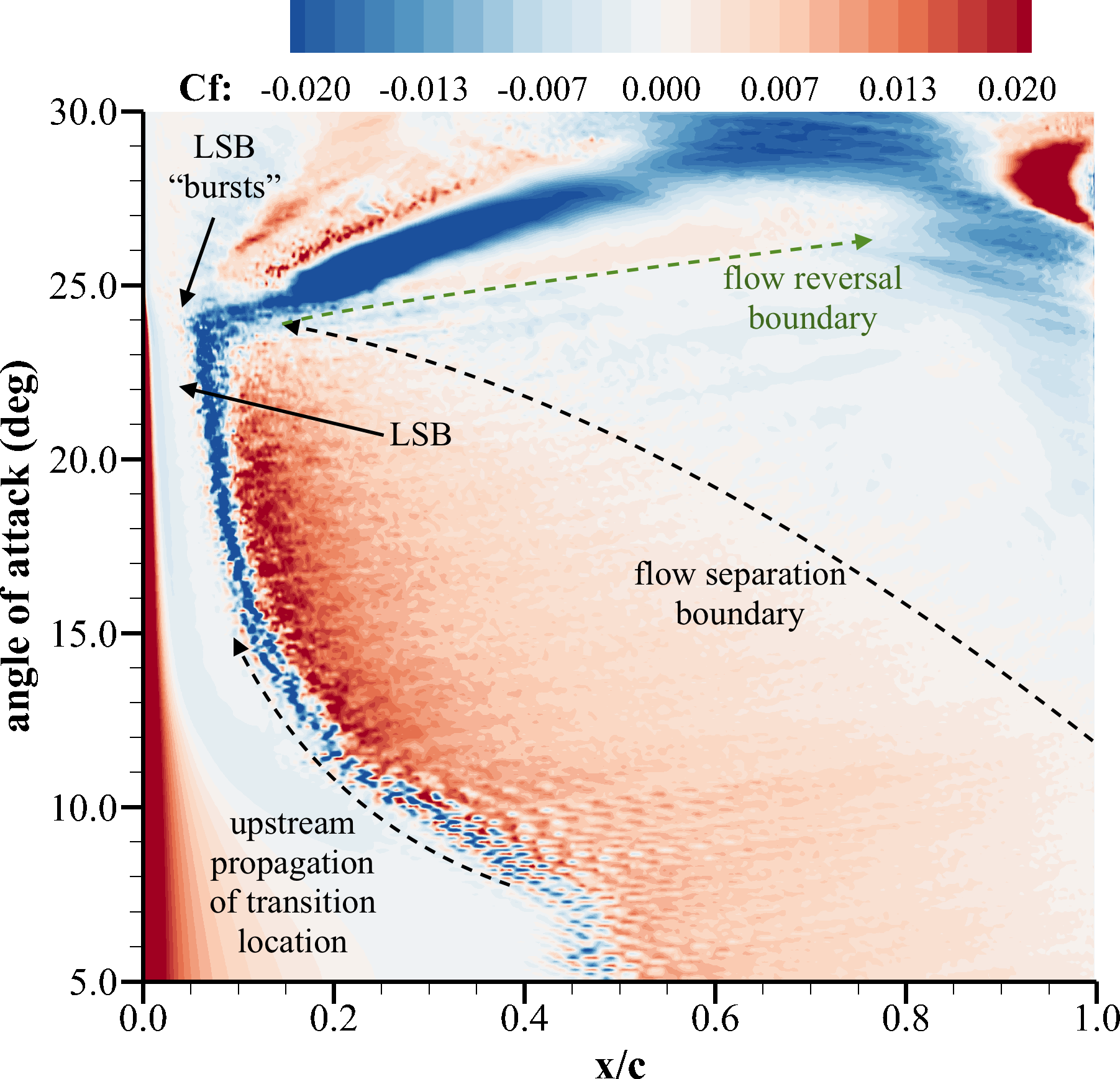}}
  \caption{Contours of span-averaged skin friction coefficient ($\langle C_f
    \rangle$) on the suction side of the four airfoils through the constant
    pitch-rate motion.}
  \label{fig:Cf_contours}
\end{figure}

We investigate the mechanism of stall onset for the cases considered here by
analyzing the details of the flowfield over the suction surface for each
airfoil. Figures~\ref{fig:Cp_contours} and~\ref{fig:Cf_contours} respectively
plot spanwise averaged contours of $-C_P$ and $C_f$ (denoted by -$\langle C_P
\rangle$ and $\langle C_f \rangle$ respectively) on the suction side of the
airfoil as functions of chordwise distance and angle of attack, $\alpha$. This
representation is similar to $x-t$ diagrams with $\alpha$ representing time
($t$) scaled by the pitch rate (since the pitch rate is constant). $x-t$
diagrams are useful to identify characteristics of hyperbolic equations.
Contour plots are shown for all four cases. The sequence of flow events
identified earlier in Section~\ref{sec:dynamic_results} are clearly seen in the
contour plots. The transition location is identified by the boundary where the
2D instability modes (seen clearly in Fig.~\ref{fig:Cf_contours} as alternating
blue and red spots) start to appear. The transition location moves upstream
with increasing $\alpha$. The speed at which the transition location moves
upstream reduces with increasing airfoil thickness. The LSB forms near the
leading edge (marked by leveling off of chordwise variation of $\langle C_P
\rangle$) and is sustained up to approximately $\alpha =
11^\circ,~15^\circ,~19^\circ$, and $23^\circ$ for the 9\%,12\%, 15\%, and 18\%
thick airfoils respectively.

\begin{figure}[htb!]
  \centering
  \subfloat[Variation of $\langle C_P \rangle$ with arc length on the suction side before LSB burst]{\incfig[width=0.45\columnwidth]{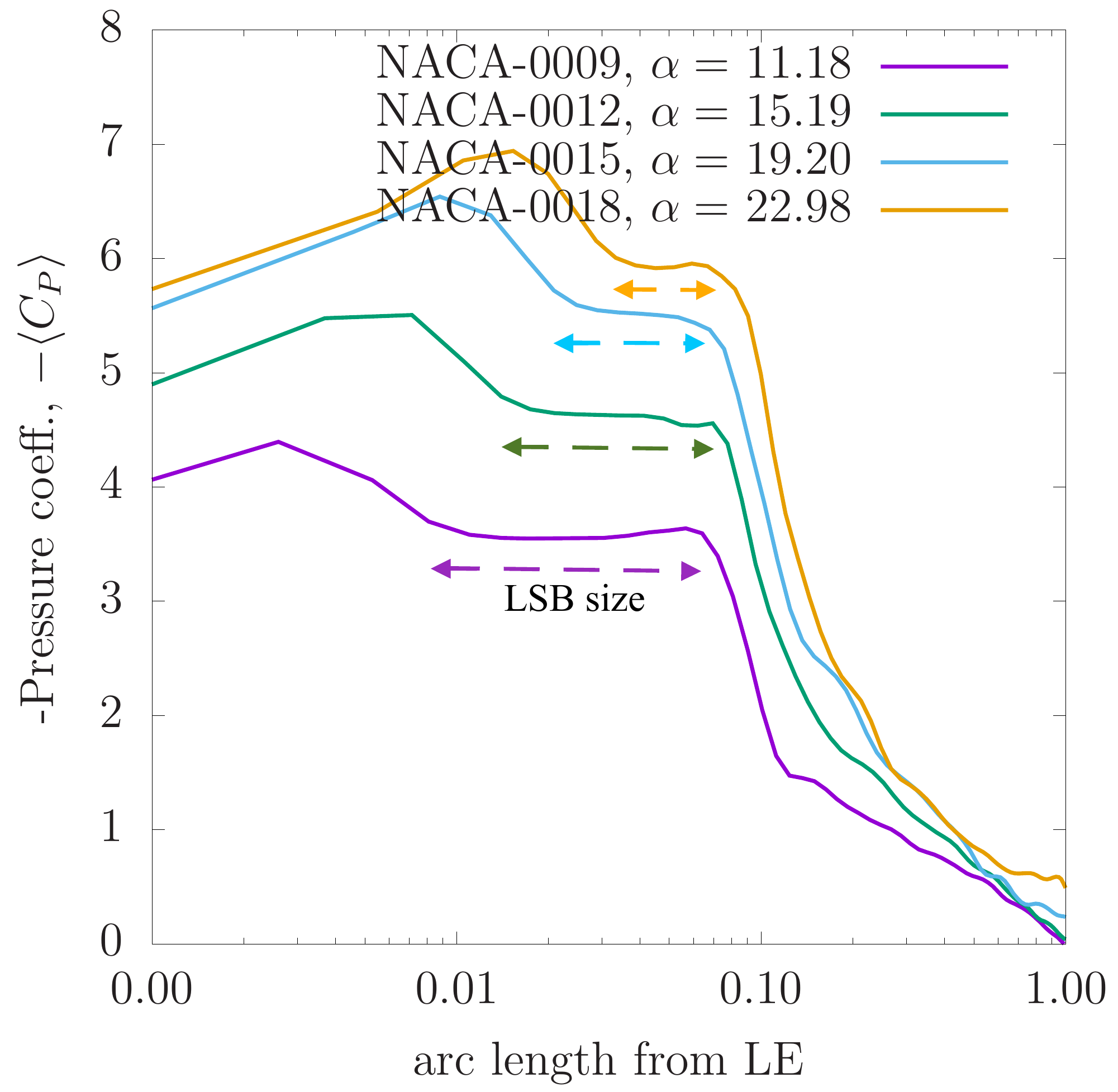}} \qquad
  \subfloat[$\langle C_P \rangle$ variation with $\alpha$ at $x/c=0.005$]                           {\incfig[width=0.45\columnwidth]{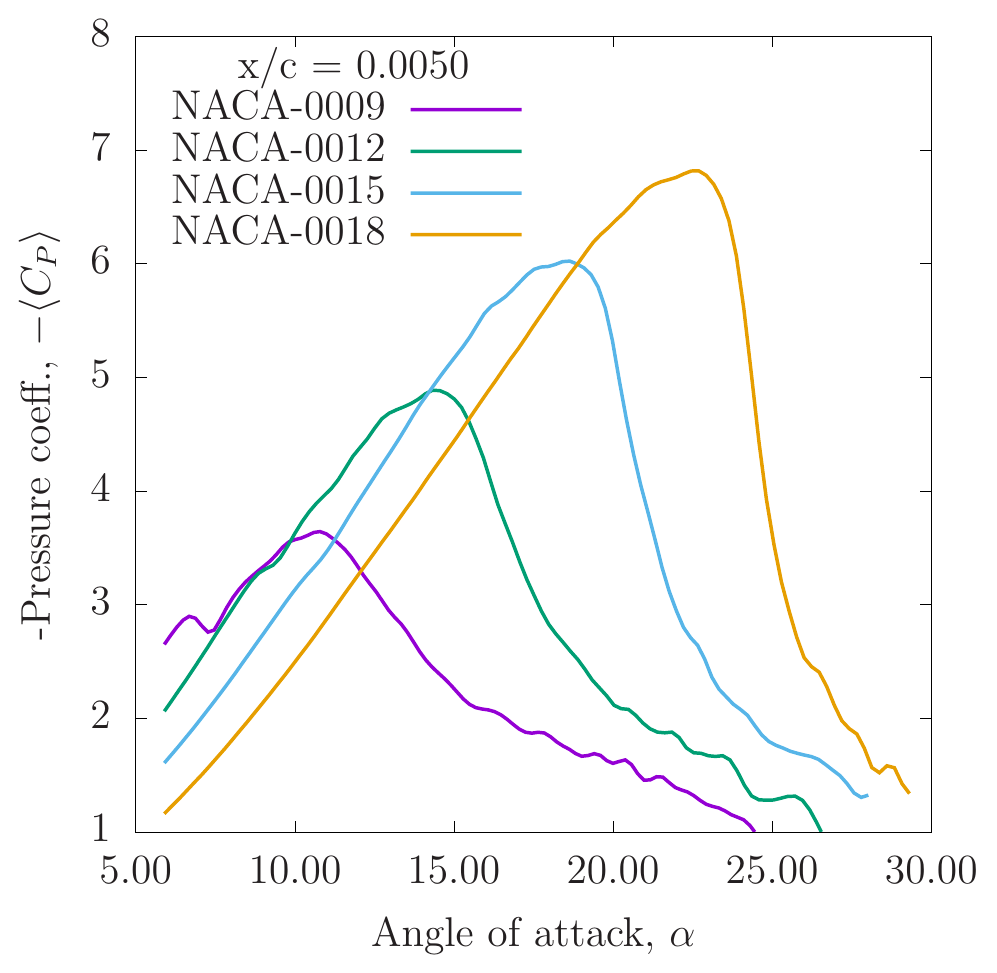}}
  \caption{Span-averaged aerodynamic pressure coefficient (-$\langle C_P
    \rangle$) variation: (a) with arc length measured from the airfoil leading
    edge just before the LSB bursts, and (2) with angle of attack at $x/c=0.005$ as
    each airfoil is pitched up at a constant rate.} 
  \label{fig:LSB_Cp}
\end{figure}

Figure~\ref{fig:LSB_Cp} (a) plots the variation of $C_P$ with arc length
measured from the leading edge for each airfoil just before the LSB collapses.
The abscissa is plotted on a logarithmic scale to zoom in on the LSB. The size
of the LSB is clearly seen to reduce with airfoil thickness. It is also
observed that the thickest airfoil (NACA-0018) experiences the largest increase
in peak $-\langle C_P \rangle$, quite in contrast with integrated lift increase due to dynamic
stall, which is observed to be highest for NACA-0009 (see
Fig.~\ref{fig:dyn_ClCdCm} (a)). This is due to larger leading edge radius of
curvature in thicker airfoils which alleviates the increase in adverse pressure
gradient due to airfoil pitch up motion, hence sustaining the LSB to higher
$\alpha$. A similar observation has been reported in
Ramesh~\etal~\cite{ramesh2011augmentation}, which defines a leading edge
suction parameter (LESP) and identifies the critical value of LESP for a given
airfoil geometry at which the flow separates at the leading edge. The LESP is
defined in an inviscid sense as the flow velocity at the leading edge of the
airfoil; a viscous equivalent of LESP would be static pressure with opposite
sign. Ramesh~\etal~\cite{ramesh2011augmentation} remark that the critical LESP
should increase with increasing airfoil thickness.

The LSB ``burst'' is marked by a sudden loss in suction near the leading edge
with increasing $\alpha$. Figure~\ref{fig:LSB_Cp} (b) plots the variation with
$\alpha$ of $-\langle C_P \rangle$ on the suction side of each airfoil at
$x/c=0.005$. The suction pressure peak collapse is more sudden for the thicker
airfoils.  The collapse of the suction peak is followed immediately by the
formation of the dynamic stall vortex (DSV). These events are notated in the
plots in Figs.~\ref{fig:Cp_contours} and~\ref{fig:Cf_contours}. The locus of
the DSV is clearly visible in Fig.~\ref{fig:Cp_contours} as a hotspot streak
running from left to right at an angle (marked with a blue arrow); the angle
determined by the speed at which the DSV convects along the airfoil chord, and
the color intensity signifying the additional suction induced by the DSV. The
{\em chordwise} convection speeds of the DSVs, computed using the slopes of the
hotspot streaks, are: 0.15, 0.18, 0.24, and 0.30 for the 9\%,12\%, 15\%, and
18\% thick airfoils respectively.  Note that the freestream flow speed is 1.0.
The apparent increase in convection speed with airfoil thickness is due to the
fact that the DSV formation and propagation occur at higher pitch angles with
increasing thickness. This is because, at higher pitch angles, the flow speed
over the entire airfoil is higher for thicker airfoil corresponding to the
higher suction ($-\langle C_P \rangle$) seen in Fig.~\ref{fig:LSB_Cp} (a) for
thicker airfoils.  A small contribution to the difference in chordwise
convection speed of the DSV also arises from the following. The DSV does not
actually convect along the airfoil chord; it moves approximately in the
direction of the freestream velocity vector. The DSV convection speed measured
using the slopes of the hot streaks in Fig.~\ref{fig:Cp_contours} is the
projection of the actual speed onto the direction of the chord line. Since the
airfoil pitch angle at the point when the DSV forms increases with airfoil
thickness, the projected chordwise convection speed would be higher for thicker
airfoils even if the actual (physical) convection speeds are the same.

Flow reversal on the airfoil suction surface is investigated to find out if it
plays a role in dynamic stall onset.  Region of flow reversal are identified in
Fig.~\ref{fig:Cf_contours} by negative values of $\langle C_f \rangle$. A
two-color scheme is chosen for the contour plots in Fig.~\ref{fig:Cf_contours}
to aid in visually identifying the reverse-flow regions. It is seen that for
NACA-0009, there is virtually no flow reversal near the trailing edge by the
time the DSV forms and stall occurs. In the NACA-0012 case, there is a hint of
flow reversal (faint blue contours between $12^\circ < \alpha < 18^\circ$;
region between the dashed black and green lines in Fig.~\ref{fig:Cf_contours}
(b)) localized near the trailing edge. The NACA-0015 case however shows a
moderate size flow separation region that reaches almost up to 30\% chord when
the LSB bursts and dynamic stall begins. In these three cases, the dynamic
stall onset is clearly triggered by the bursting of the LSB and hence can be
categorized as leading edge stall. For the thickest airfoil tested (NACA-0018)
however, the flow reverse flow region in the turbulent boundary layer reaches
the location of the LSB ($x/c \sim 0.18$) exactly at the time when the LSB
collapses. In this case, it is difficult to isolate the mechanism that triggers
dynamic stall. The trailing edge separation region interacting with the LSB
could be the mechanism that causes the airfoil to stall.

\begin{figure}[htb!]
  \centering
  \incfig[width=0.7\columnwidth]{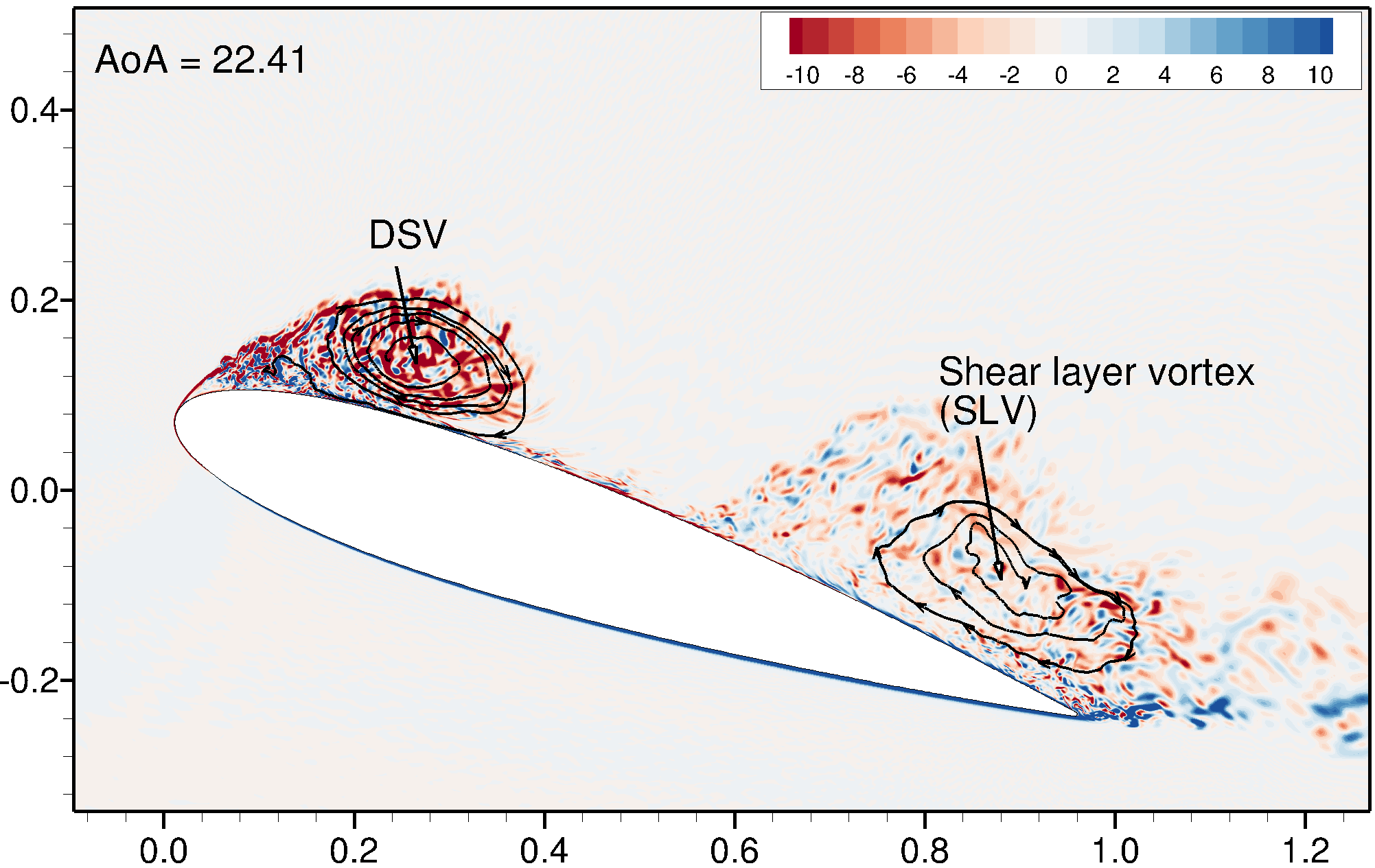}
  \caption{Vorticity contours for NACA-0015 airfoil at $\alpha=22.41^\circ$ identifying
  the shear layer vortex (SLV) and the DSV.}
  \label{fig:SLV}
\end{figure}

Another characteristic, that is readily observed in Fig.~\ref{fig:Cf_contours}
(c), is the left-to-right running line that starts at the LSB-burst location
and convects at a speed greater than that of the DSV (shallower angle in the
plot). This characteristic is denoted by the green dashed line with an
arrowhead in the figure. The $\langle C_f \rangle$ changes sign across this characteristic -
from negative to positive as $\alpha$ is increased. A moderate drop in suction
pressure is also observed across this characteristic
(Fig.~\ref{fig:Cp_contours}). As the DSV grows, some of the viscous boundary
layer vorticity rolls up into it. The remaining vorticity rolls up further
downstream into a shear layer vortex (SLV). The DSV and the LSV are visualized
in Fig.~\ref{fig:SLV} using vorticity contours and streamlines. In between the
DSV and the LSV, there is a region of positive $\langle C_f \rangle$ due to the interplay
between the freestream and the velocity induced by the DSV. This is also seen
in Fig.~\ref{fig:Qcrit_0012} (e \& f) where the region between the DSV and the
LSV shows turbulent eddies stretched in the streamwise direction due to the
flow locally accelerated by the DSV. The characteristic referred to above,
marks the trailing end of the SLV. The propagation speed of this characteristic
is nearly equal to unity as the SLV convects with the local flow speed along
the chord.

\begin{figure}[htb!]
  \subfloat[NACA 0009]{\incfig[width=0.45\columnwidth]{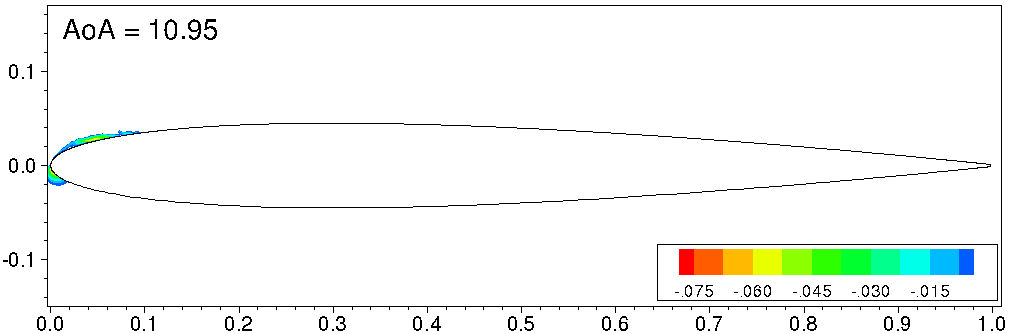}} \qquad
  \subfloat[NACA 0012]{\incfig[width=0.45\columnwidth]{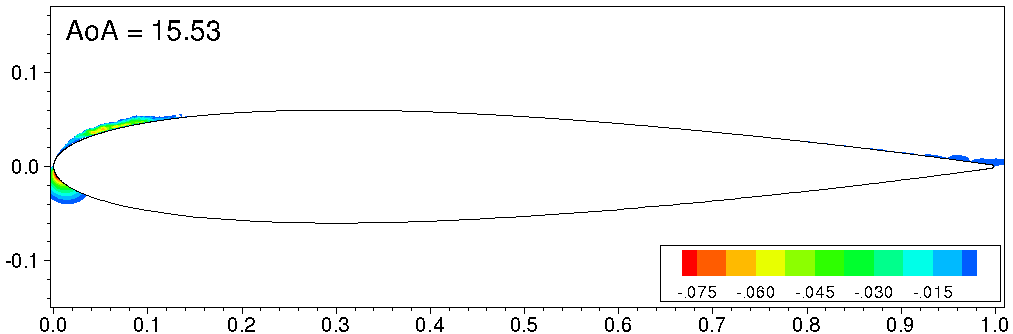}} \\
  \subfloat[NACA 0015]{\incfig[width=0.45\columnwidth]{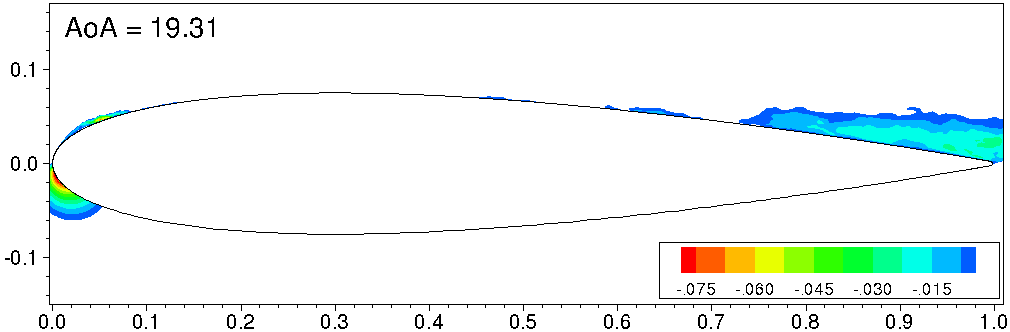}} \qquad
  \subfloat[NACA 0018]{\incfig[width=0.45\columnwidth]{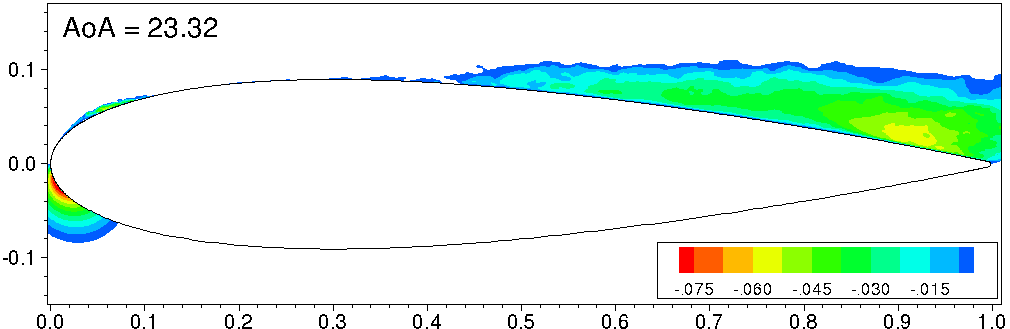}} \\
  \caption{Contours of blade-relative chord-wise flow velocity for the four
  airfoils immediately before onset of dynamic stall. The contours are cut-off
  above 0 to identify reverse flow regions.}
  \label{fig:flow_reversal}
\end{figure}

Figure~\ref{fig:flow_reversal} plots instantaneous contours of chordwise blade
relative velocity for each airfoil immediately prior to onset of dynamic stall.
The contours are cutoff above the zero value to show only the reverse flow
regions. Reverse flow region is clearly visible in the aft portion of the
relatively thick airfoils (NACA-0015 and NACA-0018), while the 9\% and 12\%
thick airfoils show almost no flow reversal. While these plots provided a good
qualitative view of how far upstream the reverse flow region reaches at the
onset of dynamic stall, the skin friction coefficient is examined next for a
quantitative assessment.

\begin{figure}[htb!]
  \subfloat[-Span-averaged pressure coeff., $-\langle C_P \rangle$]    {\incfig[width=0.45\columnwidth]{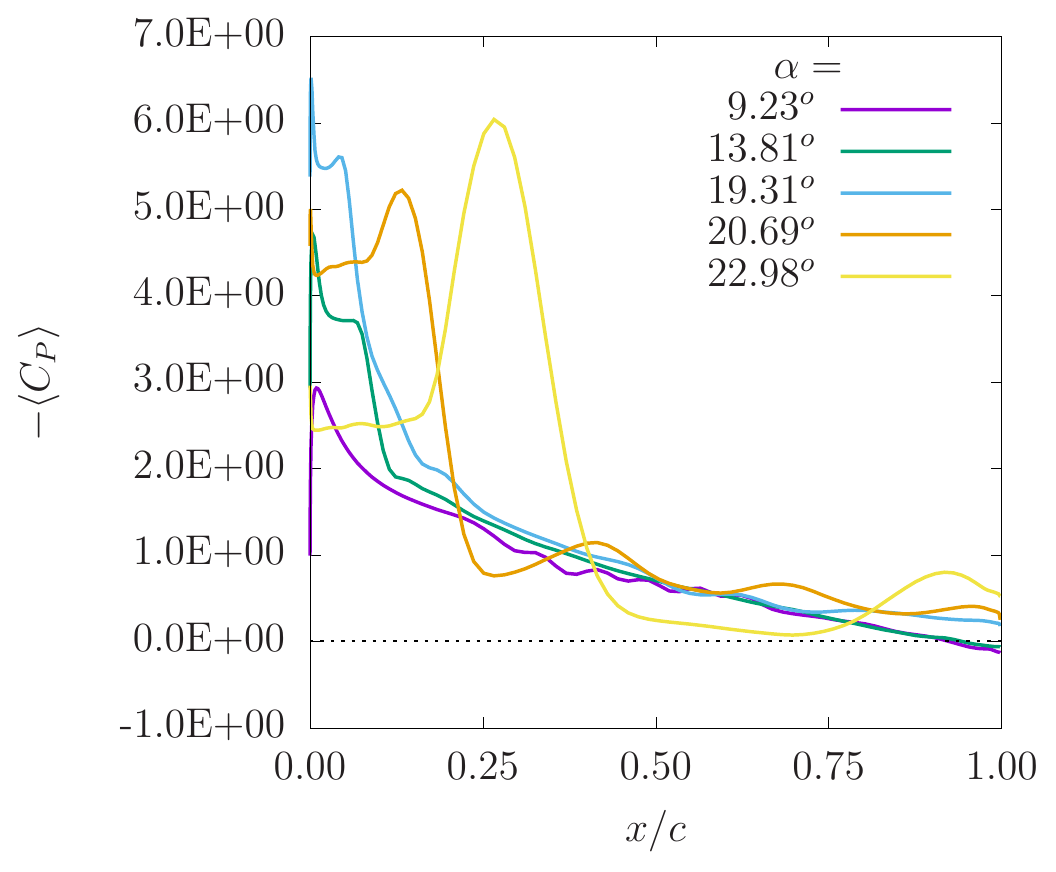}} \qquad
  \subfloat[Span-averaged skin friction coeff., $\langle C_f \rangle$] {\incfig[width=0.45\columnwidth]{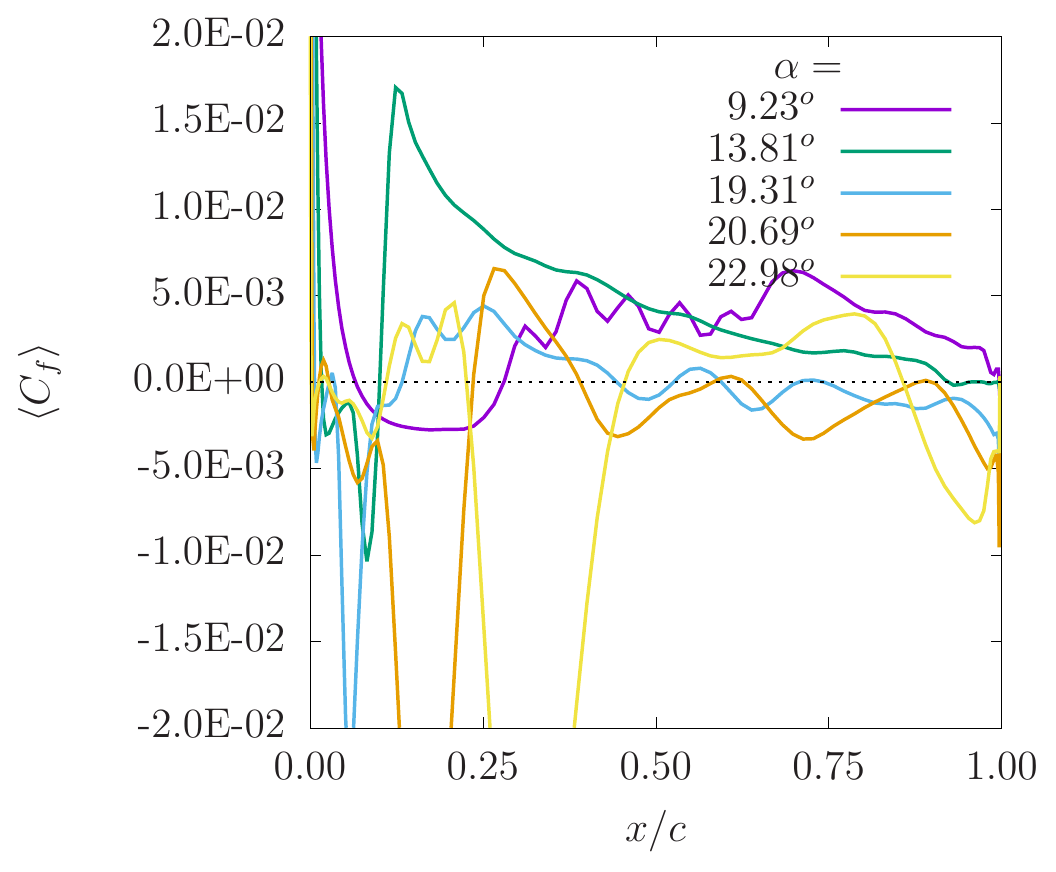}} \\
  \caption{Distributions of $-\langle C_P \rangle$ and $\langle C_f \rangle$
    along the NACA-0015 chord at five angles of attack during the pitch up
    maneuver.}
  \label{fig:NACA-0015_CpCf}
\end{figure}

Figure~\ref{fig:NACA-0015_CpCf} shows line plots of $-\langle C_P \rangle$ and
$\langle C_f \rangle$ along the NACA-0015 airfoil chord at five different
angles of attack ($\alpha$) during the pitch-up maneuver. The $\alpha$ values
are selected to illustrate a few interesting stages in the pitch up maneuver.
At $\alpha=9.23^\circ$, the laminar boundary layer over the airfoil locally
separates (see $\langle C_f \rangle$ plot) and transitions; the transition
region shows oscillations corresponding to the instability modes in both
$\langle C_P \rangle$ and $\langle C_f \rangle$. At $\alpha=13.81^\circ$, the LSB
is securely positioned close to the airfoil leading edge and the boundary layer
transitions abruptly right behind the LSB. Some evidence of the turbulent
boundary layer separating near the trailing edge is also visible. Further
increase in $\alpha$ to $19.31^\circ$ causes the LSB to move upstream and shrink in
size. At this time, the turbulent boundary layer is separated beyond mid-chord
($\langle C_f \rangle <0$). The LSB bursts as $\alpha$ is increased beyond
$19.31^\circ$ and the DSV forms. The DSV is seen as locally increased $C_P$ value
in the curves for $\alpha=20.69^\circ$ and $22.98^\circ$. As the DSV forms and convects
downstream, some part of the turbulent boundary layer reattaches (as seen in
the $C_f$ curve for $\alpha=22.98^\circ$) due to the large induced velocity by the
DSV. This is marked as ``flow reversal boundary'' in
Figs.~\ref{fig:Cp_contours} and~\ref{fig:Cf_contours}.

\begin{figure}[htb!]
  \incfig[width=0.6\columnwidth]{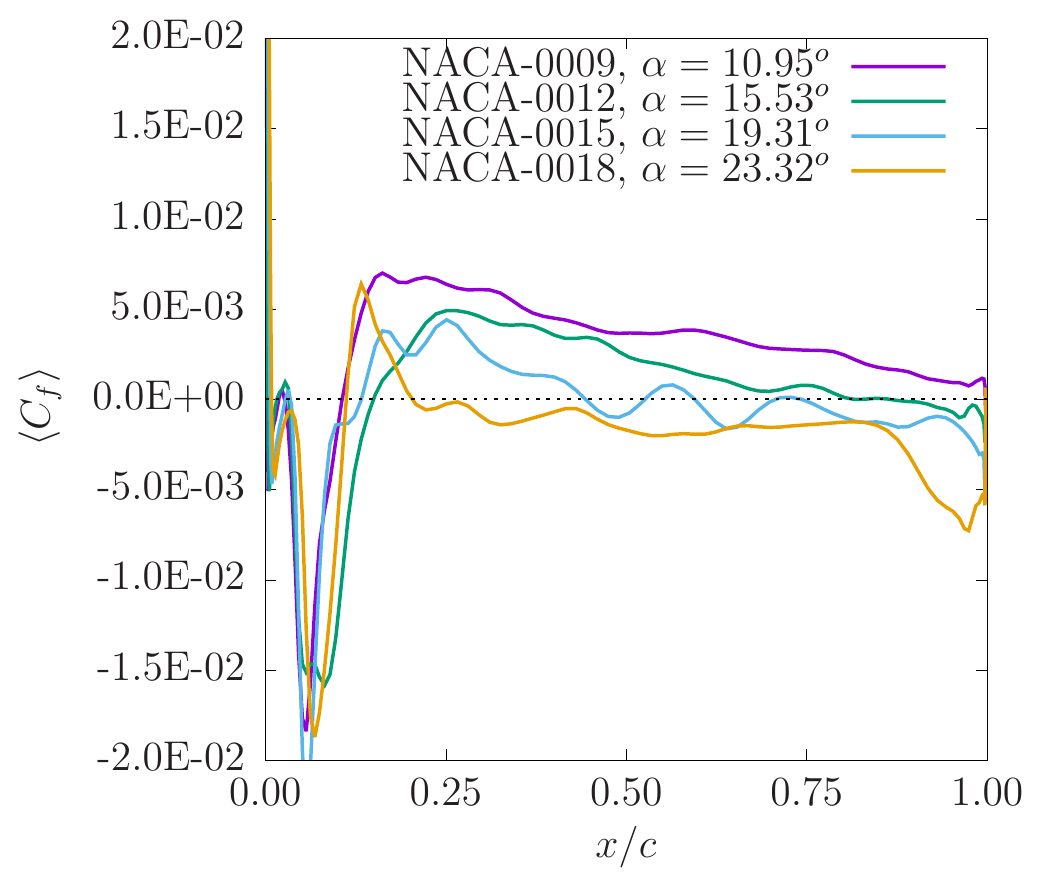}
  \caption{$\langle C_f \rangle$ distributions on the suction surfaces of the
    four airfoils immediately before onset of dynamic stall.}
  \label{fig:Cf_cmp}
\end{figure}

Figure~\ref{fig:Cf_cmp} compares $C_f$ distributions between the four airfoils
taken immediately prior to the bursting of the LSB. No flow separation is seen
near the trailing edge for the thinnest airfoil. The NACA-0012 simulation shows
reverse flow in a very small region near the trailing edge. More than 50\% of
the NACA-0015 airfoil experiences reverse flow before LSB burst, while for
NACA-0018, the turbulent flow separation point reaches the edge of the LSB
before onset of stall. The close proximity of the turbulent flow separation
with the LSB suggests that the stall onset could be caused either by the
bursting of the LSB or by the separated turbulent boundary layer interacting
with the LSB for the NACA-0018 airfoil.

%%%%%%%%%%%%%%%%%%%%%%%%
\section{Conclusions}
\label{sec:conclusions}
%%%%%%%%%%%%%%%%%%%%%%%%%%%%%%%%%%%%%%%%%%%%
Onset of dynamic stall is investigated at $Re_c=2\times10^5$ for four symmetric
NACA airfoils of varying thickness -  9\%, 12\%, 15\%, and 18\%. A constant
rate pitch-up airfoil motion about the quarter-chord point is investigated
using wall-resolved large eddy simulations. Comparisons are drawn against XFOIL
for static simulations at angle of attack, $\alpha=4^\circ$. Overall, the
agreement between FDL3DI and XFOIL in predicting $C_P$ and $C_f$ distributions
is quite good. XFOIL however does not capture the two-stage transition process
observed in FDL3DI for relatively thinner (9\% and 12\%) airfoils. XFOIL also
does not show any significant change in $\partial C_f/\partial x$ with airfoil
thickness, whereas FDL3DI predicts a large increase with thickness.

The effect of finite span size is evaluated by investigating spanwise coherence
of pressure. It is found that while the solution is highly correlated along the
entire span in the post-stall region, the correlation is rather small in the
stall incipience region and hence onset of stall can be investigated with the
span length of 10\% chord utilized in this study.

Dynamic simulations show the following sequence of events: (1) upstream
movement of the transition location, (2) formation of a laminar separation
bubble (LSB) and rise in suction peak pressure, (3) LSB burst followed by
formation of the dynamic stall vortex (DSV), (4) roll up of boundary layer
vorticity into a vortex (shear layer vortex or SLV), (5) sharp increase in
pitch-down moment (moment stall), and (5) precipitous drop in airfoil lift
(lift stall). While all the airfoils undergo the same sequence of events, the
duration of each event and the associated aerodynamics differ substantially
with airfoil thickness. The thinnest airfoil tested (NACA-0009) experiences the
largest increase in sectional lift coefficient whereas the highest peak suction
pressure is obtained for the thickest airfoil (NACA-0018). 

Comparisons of ${C_P}_{rms}$, where mean $C_P$ is obtained via low-pass
filtering the solution, show high correlation between increase in ${C_P}_{rms}$
and sharp increase in $C_f$, thus verifying that ${C_P}_{rms}$ measurements can
be effectively used to locate boundary layer transition.

Spatio-temporal diagrams of span-averaged $-C_P$ and $C_f$ clearly show the
different stages of dynamic stall, and highlight the differences between the
different airfoils. The $\alpha$ up to which the LSB is sustained increases
with airfoil thickness. The peak value of $-C_P$ near airfoil leading edge,
increases with airfoil thickness. In all cases, the LSB bursts is followed by
the formation of the DSV, however the characteristics of the DSV and its
convection speed vary with airfoil thickness, with the highest speed for the
thickest airfoil.

Investigation of skin friction coefficient on the suction surface shows that
while turbulent boundary layer separation is nearly non-existent for NACA-0009,
the separation (flow reversal) region for NACA-0018 extends from the trailing
edge all the way up to the LSB location immediately before dynamic stall
occurs. This observation suggests that stall onset could have been triggered by
the turbulent separation region reaching up to and interacting with the LSB for
NACA-0018, and the possibility that mechanism of stall onset gradually changes
with airfoil thickness from that due solely to LSB burst to that due to
interaction of trailing edge separation with the LSB.

\begin{acknowledgments}
\label{sec:acknowledgement}
Funding for this research is provided by the AFOSR Summer Faculty Fellowship
program and by the National Science Foundation under grant number NSF/
CBET-1554196. Computational resources are provided by NSF XSEDE (Grant
\#TG-CTS130004) and the Argonne Leadership Computing Facility, which is a DOE
Office of Science User Facility supported under Contract DE-AC02-06CH11357. 

The second author would like to acknowledge support by AFOSR under a task
monitored by Dr. D. Smith, and by a grant of HPC time from the DoD HPC Shared
Resource Centers at AFRL and ERDC. Technical support for the FDL3DI software
provided by Dr. Daniel Garmann of the Air Force Research Laboratory is
acknowledged.
\end{acknowledgments}

% You should use BibTeX and apsrev.bst for references
% Choosing a journal automatically selects the correct APS
% BibTeX style file (bst file), so only uncomment the line
% below if necessary.

%\section*{References}
%\bibliographystyle{apsrev}
\bibliographystyle{aiaa}
\bibliography{references}

\end{document}